\documentclass{aastex63}

\usepackage{graphicx}
\usepackage{amssymb}
\usepackage{amsmath}
\usepackage{epstopdf}
\usepackage{gensymb}
\usepackage{natbib}
\usepackage{appendix}
\usepackage{tabu}
\usepackage{color}
\newcommand{\sref}[1]{Section~\ref{#1}}
\newcommand{\fref}[1]{Figure~\ref{#1}}
\newcommand{\eqn}[1]{Equation~(\ref{#1})}

\newcommand{\gal}{\mbox{Gaia-G}}
\newcommand{\kep}{\mbox{\it Kepler}}
\newcommand{\sj}{\mbox{Small-JASMINE}}
\defcitealias{Sasha2021}{Paper I}
\defcitealias{Nina1}{N20}

\begin{document}
\title{Predictions of Astrometric Jitter for Sun-like Stars. II. Dependence on Inclination, Metallicity, and Active-Region Nesting}
\accepted{July 2, 2021}
\shorttitle{Activity-induced astrometric jitter}
\shortauthors{Sowmya et al.}

\correspondingauthor{K.~Sowmya}
\email{krishnamurthy@mps.mpg.de}

\author[0000-0002-3243-1230]{K.~Sowmya}
\affiliation{Max-Planck-Institut f\"ur Sonnensystemforschung, Justus-von-Liebig-Weg 3, 37077 G\"ottingen, Germany}

\author[0000-0001-6090-1247]{N.-E.~N\`emec}
\affiliation{Max-Planck-Institut f\"ur Sonnensystemforschung, Justus-von-Liebig-Weg 3, 37077 G\"ottingen, Germany}

\author[0000-0002-8842-5403]{A.~I.~Shapiro}
\affiliation{Max-Planck-Institut f\"ur Sonnensystemforschung, Justus-von-Liebig-Weg 3, 37077 G\"ottingen, Germany}

\author[0000-0001-6163-0653]{E.~I\c{s}{\i}k}
\affiliation{Department of Computer Science, Turkish-German University,
\c{S}ahinkaya Cd. 94, Beykoz, 34820 Istanbul, Turkey}

\author[0000-0002-0929-1612]{V.~Witzke}
\affiliation{Max-Planck-Institut f\"ur Sonnensystemforschung, Justus-von-Liebig-Weg 3, 37077 G\"ottingen, Germany}

\author[0000-0002-8440-1455]{A.~Mints}
\affiliation{Leibniz-Institut f\"ur Astrophysik Potsdam, 14482 Potsdam, Germany}

\author[0000-0002-1377-3067]{N.~A.~Krivova}
\affiliation{Max-Planck-Institut f\"ur Sonnensystemforschung, Justus-von-Liebig-Weg 3, 37077 G\"ottingen, Germany}

\author[0000-0002-3418-8449]{S.~K.~Solanki}
\affiliation{Max-Planck-Institut f\"ur Sonnensystemforschung, Justus-von-Liebig-Weg 3, 37077 G\"ottingen, Germany}
\affiliation{School of Space Research, Kyung Hee University, Yongin, Gyeonggi 446--701, Korea}

\begin{abstract}
Ultra-precise astrometry from the Gaia mission is expected to lead to astrometric detections of more than 20,000 exoplanets in our Galaxy. One of the factors that could hamper such detections is the astrometric jitter caused by the magnetic activity of the planet host stars. In our previous study, we modeled astrometric jitter for the Sun observed equator-on. In this work, we generalize our model and calculate the photocenter jitter as it would be measured by the Gaia and \sj{} missions for stars with solar rotation rate and effective temperature, but with various values of the inclination angle of the stellar rotation axis. In addition, we consider the effect of metallicity and of nesting of active regions (i.e. the tendency of active regions to emerge in the vicinity of each other). We find that, while the jitter of stars observed equator-on does not have any long-term trends and can be easily filtered out, the photocenters of stars observed out of their equatorial planes experience systematic shifts over the course of the activity cycle. Such trends allow the jitter to be detected with continuous measurements, in which case it can interfere with planet detectability. An increase in the metallicity is found to increase the jitter caused by stellar activity. Active-region nesting can further enhance the peak-to-peak amplitude of the photocenter jitter to a level that could be detected by Gaia.
\end{abstract}


\section{Introduction}
\label{sec:intro}
Over 4300 exoplanets have so far been discovered in our Galaxy\footnote{\url{http://exoplanet.eu/catalog}}. The majority of these discoveries have been made with the radial velocity and transit photometry methods. Using ground-based observations of periodic variations in stellar radial velocity caused by the gravitational interaction with a planet, \citet{1995Natur.378..355M} discovered the first exoplanet orbiting a Sun-like star. This commenced many dedicated planet-search programs. For example, the High Accuracy Radial velocity Planet Searcher \citep[HARPS;][]{2000SPIE.4008..582P} spectrograph discovered over 130 exoplanets, making it the most successful planet finder to date that utilizes the radial velocity technique. The Convection, Rotation and planetary Transit \citep[CoRoT;][]{2003A&A...405.1137B,2006ISSIR...6..221M} became the first space-based mission to discover exoplanets using transit photometry. The \kep{} mission \citep{2010Sci...327..977B}, which used the same method as CoRoT, has so far been the most successful planet hunting mission, with its discoveries making up for more than half the number of known exoplanets in our Galaxy. The ongoing survey by the Transiting Exoplanet Survey Satellite \citep[TESS;][]{2014SPIE.9143E..20R} has already made significant exoplanet discoveries and further discoveries are anticipated from the upcoming PLAnetary Transits and Oscillations of stars \citep[PLATO;][]{2014ExA....38..249R} space mission.

One of the other promising methods for finding exoplanets has astrometry as its basis. The astrometric technique requires precise measurement of the minuscule `stellar wobble' caused by the planetary companion (i.e. changes in the position of a star introduced by the revolution of a star around the star-planet barycenter). Owing to the absence of such extremely precise measurements, only a few astrometric detections have been reported to date (e.g., the catalog of exoplanets at \url{http://exoplanet.eu/catalog} lists only 12 detections with astrometry). \mbox{HD 176051b} was the first exoplanet to be discovered using astrometry \citep{2010AJ....140.1657M} and the latest discovery, \mbox{TVLM 513-46546b}, was reported by \citet{2020AJ....160...97C}. However, a drastic change in the situation is imminent with the ultra-precise astrometric capability of the Gaia space observatory \citep{2016A&A...595A...1G}, which was launched in December 2013. In particular, it is estimated that for a mission duration of 10 years, data from Gaia would lead to a discovery of some 70,000 exoplanets \citep{2014ApJ...797...14P}. The Japanese mission \sj{} \citep{2013IAUS..289..433Y,2014SPIE.9143E..0ZU}, planned for launch in 2024, will complement Gaia's exoplanet detections by carrying out very precise astrometric measurements in the near-infrared.

The wobbling of a star around the star-planet barycenter leads to a displacement of the stellar photocenter. Another potential source of photocenter excursions of a star is its magnetic activity \citep[see,  e.g.,][]{2008NewA...13...77L}. Surface magnetic features, such as spots and faculae, not only lead to photometric variability and radial velocity changes, but also induce astrometric jitter (i.e. the displacement of the photocenter). Dark starspots cause a decline in the surface brightness and repel the photocenter, whereas the bright faculae increase the surface brightness attracting the photocenter towards them. The photocenter displacement caused by the surface magnetic activity is time-dependent due to the evolution of magnetic features and stellar rotation. All in all, the magnetically-driven jitter is a source of noise in the astrometric data and may pose a limitation on the determination of stellar parallax and proper motion in addition to the detection and characterization of exoplanets. 

In view of the Space Interferometry Mission (which was cancelled in late 2010) and Gaia, several studies estimated the level of the jitter the Sun would have \citep[e.g.][]{2007A&A...476.1389E,2008SPIE.7013E..2KC,2009ApJ...707L..73M, 2010A&A...512A..39M,2011A&A...528L...9L}. It was found that for the Sun observed from the ecliptic plane the jitter could be comparable to the signal introduced by the Earth orbiting the Sun (0.3\,$\mu$as for the Sun located 10 pc away from the observer).  The second release of Gaia astrometric data \citep{2018A&A...616A...2L} has led to a renewed interest in exploring the impact of starspot jitter on the astrometric signal from a planet. Recently, \citet{2018MNRAS.476.5408M} developed a simple model to show that the precision of Gaia astrometry is insufficient to detect starspot-induced jitter from stars with near-solar activity levels, though it is sufficient to detect jitter of nearby active stars. \citet{2019A&A...627A..56M} proposed a model to simulate astrometric time series for solar-type stars. Using these simulated time series, \citet{2020A&A...644A..77M} quantified the effect of stellar activity on astrometric measurements, and its impact on the search for Earth-mass planets in the habitable zones of old main-sequence solar-type stars.

\citet[][hereafter Paper I]{Sasha2021} developed a model, that employs the observed distribution of spots and faculae on the visible solar disk, to calculate the astrometric jitter of the Sun. It is an extension of the Spectral And Total Irradiance REconstruction (SATIRE) model \citep{2000A&A...353..380F,2003A&A...399L...1K}, which is one of the most successful models for reconstructing solar irradiance variations. They found that the peak-to-peak amplitude of activity-induced solar jitter in the \gal{} passband can reach 0.5\,$\mu$as for the Sun located 10 pc away from the observer. They also showed that, by dint of the peculiar center to limb variation of their brightness contrast, solar faculae cause a displacement comparable to that brought about by spots, though in opposite direction analogous to irradiance variations.

The model developed in \citetalias{Sasha2021} was limited to calculating astrometric jitter for the Sun observed from the ecliptic plane. At the same time the astrometric jitter is expected to depend on the inclination (i.e. the angle between the stellar rotation axis and the direction to the observer), the metallicity, as well as the degree of active-region nesting (i.e. the tendency of active regions to emerge in the vicinity of recent magnetic flux emergence). The stellar inclination influences the visibility of magnetic features \citep[see e.g.][]{2001A&A...376.1080K,2014A&A...569A..38S,Nina1}. The metallicity governs the contrasts of the magnetic features with respect to the surrounding quiet regions \citep[e.g.][]{Veronika_2018, Veronika_2020}. Active-region nesting affects the spatial distribution of magnetic features \citep[e.g.][]{1986SoPh..105..237C,Emre2020} and is known to have a strong effect on the photometric variability \citep{Emre2020}. Through its direct effect on the surface distribution of brightness, active-region nesting is likely to have a significant influence on the astrometric jitter. Therefore, in this study we extend the model presented in \citetalias{Sasha2021} to stars observed at arbitrary inclinations, with different metallicity values and active-region nesting degrees. In \sref{sec:model}, we describe the extended model used for calculating the photocenter displacements. The effects of inclination, metallicity and active-region nesting on the astrometric jitter are demonstrated in \sref{sec:res}. The test performed to check the detectability of astrometric jitter in the \gal{} passband is illustrated in \sref{sec:det}. Our conclusions are presented in \sref{sec:concl}.

\section{Modeling approach}
\label{sec:model}
The astrometric jitter is a consequence of spatial symmetry breaking about the axis joining the visible stellar disk center and the observer. The axial symmetry is generally broken when magnetic features (spots and faculae) emerge on the visible disk. In the absence of any apparent magnetic activity, the stellar photocenter is located at the center of the visible stellar disk (the shift of the photocenter due to granulation and oscillations is expected to be very small for a main-sequence star). The emergence of spots and faculae leads to a shift in the photocenter. While the dark spots repel the photocenter, the bright faculae attract it \citepalias[see Figure~1 of][]{Sasha2021}. The location of the photocenter at any given time is determined by two factors: the distribution of spots and faculae on the visible stellar disk (which is affected by active-region nesting and which appears different depending on the inclination) and their brightness contrasts (which, in turn, depend on the position due to the center to limb variations and metallicity). Therefore, our aim is to determine how sensitive the astrometric jitter caused by the surface magnetic features is to the changes in these two factors.

The fractional area coverages of the stellar disk by the magnetic features are important ingredients of our model. These area coverages depend on the inclination. Since the available observations of the Sun, which we use as the basis, are limited to the ecliptic plane (corresponding to an inclination of roughly $90\degree$), the changes in the fractional area resulting from a change in the inclination can currently be only determined from simulations. Therefore, while \citetalias{Sasha2021} employed the SATIRE-S area fractions derived by \cite{2014A&A...570A..85Y} from the observed solar intensity images and magnetograms, we utilize the fractional area coverages from \citet[][hereafter N20]{Nina1}, who employed the surface flux transport model (SFTM) of \cite{2010ApJ...719..264C} to obtain the distribution of magnetic features on the stellar surface for any given inclination angle. The SFTM is an advective-diffusive model that simulates the passive transport of the radial magnetic field on the stellar surface. The magnetic flux is injected on the surface in the form of bipolar magnetic regions. The emergent bipolar regions evolve under the influence of differential rotation and meridional flow, as well as diffusion and the cancellation between opposite magnetic polarities. 

\citetalias{Nina1} used emergence times, latitudes, tilt-angles, and separations between the two polarities in bipolar regions from the semi-empirical sunspot-group record of \citet{2011A&A...528A..82J}. The record is available for the period $1700-2010$ and constructed so as to reflect the statistical properties of the sunspot groups in the Royal Greenwich Observatory sunspot record. \citetalias{Nina1} additionally randomized the longitudes of active region emergences to ensure that there is no asymmetry between the activity on the near- and far-side of the Sun, making their model well-suited for inclination studies. The \citetalias{Nina1} model returns simulated full-surface magnetic field maps on a latitude-longitude grid with a spatial resolution of $1\degree\times1\degree$ and at a cadence of 6 hours. The area coverages by spots and faculae are then determined from these simulated surface maps \citepalias[see Section 2.3 of][for details]{Nina1}. Although the output of the \citetalias{Nina1} model allows calculating astrometric signal for the time period $1700-2010$, for simplicity, we only consider solar cycle 22 (which is a relatively strong cycle) for this study. 

Following \citetalias{Sasha2021}, we compute the shift in the photocenter caused by magnetic activity, for a given inclination $i$ as
\begin{equation}
    \begin{pmatrix}
    \Delta{X_i}(t)\\
    \Delta{Y_i}(t)
    \end{pmatrix} 
    = \frac{\sum\limits_{j} \sum\limits_{k} \alpha_{ij}^{k}(t) \begin{pmatrix} 
    {X_j} - {X}_0\\ 
    {Y_j} - {Y}_0
    \end{pmatrix}
    (F_{j}^{k} - F_{j}^0)}{(1+\mathcal{B}_{i}(t))\sum\limits_{j}F_{j}^0}\ ,
    \label{eq:centroid}
\end{equation}
where
\begin{equation}
    \mathcal{B}_{i}(t) = \sum\limits_{j} \sum\limits_{k} \alpha_{ij}^{k}(t)(F_{j}^{k} - F_{j}^0)/\sum\limits_{j}F_{j}^0\ .
    \label{eq:brightness}
\end{equation}
Here, $\Delta{X_i}$ and $\Delta{Y_i}$ are the displacements of the photocenter in $X$ and $Y$ directions, respectively. The $X$-axis is defined by the East-West line passing through the visible disk center and the $Y$-axis is perpendicular to it. The origin of this $XY$ coordinate system always coincides with the visible disk center. For ${i}=90\degree$, the stellar rotation axis is parallel to the $Y$-axis whereas for ${i}=0\degree$ it is at the origin and points along the line of sight of the observer. 

The summation `$j$' is performed over the pixels of the surface magnetic field maps, and only those pixels which are projected onto the visible stellar disk are considered for the summation. The coordinates of each pixel on the visible disk ($X_{j}$ and $Y_{j}$) are obtained by transforming from the heliographic latitude-longitude system to the heliocentric Cartesian system \citep{2006A&A...449..791T}. $X_0$ and $Y_0$ are the coordinates of the center of the visible disk. The summation `$k$' is done over three types of magnetic features, namely, faculae, spot umbra, and spot penumbra. $\alpha_{ij}^{k}(t)$ values are fractional coverages of the ${j}^{\rm th}$ pixel by the ${k}^{\rm th}$ feature for an inclination $i$, while $F_{j}^{k}$ is the photon flux of the ${k}^{\rm th}$ feature fully covering the ${j}^{\rm th}$ pixel (with $F_{j}^0$ being the quiet-star flux): 
\begin{equation}
    F_{j}^{k} = \ {\rm cos}(\delta_{j})\ \mu_{j}\  \int_{\lambda} I^{k}(\lambda,\mu_{j})\,\frac{\phi(\lambda)}{hc/\lambda}\,d\lambda\ .
    \label{eq:flux}
\end{equation}
Here, the factor ${\rm cos}(\delta_{j})$, where $\delta_{j}$ is the latitude of the pixel, accounts for the reduction in the pixel area for increasing latitude. $\mu_{j}$ is the cosine of the heliocentric angle of the pixel, which accounts for foreshortening of the pixel. $I^{{k}}(\lambda,\mu_{j})$ is the intensity of the ${k}^{\rm th}$ feature. \citet{Veronika_2018, Veronika_2020} calculated the intensities of the quiet and magnetic features at various $\mu$ values for solar and non-solar metallicity values. Here we use their pre-calculated intensities for the quiet regions and magnetic features.

\begin{figure}[ht!]
    \centering
    \includegraphics[scale=0.55]{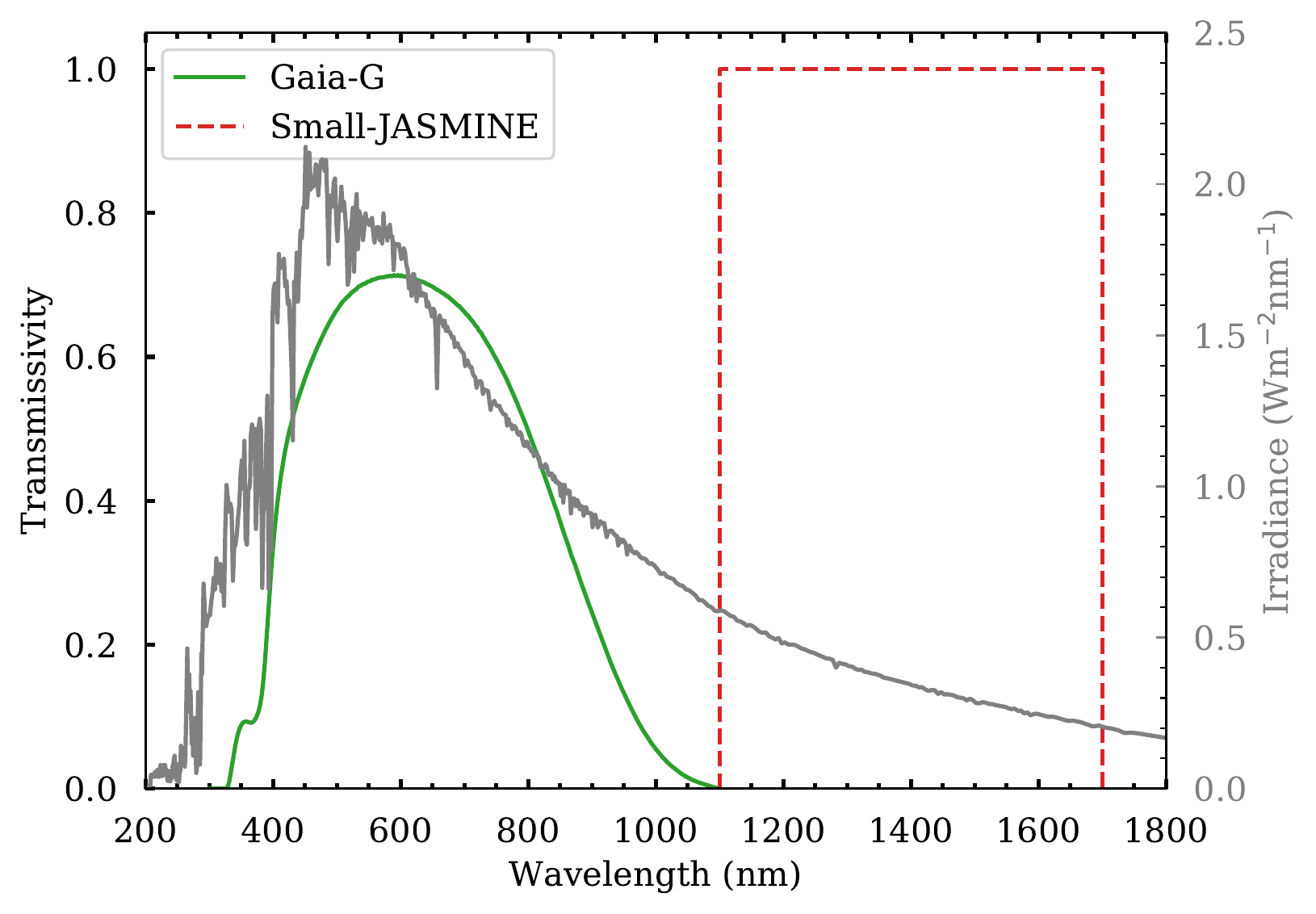}
    \caption{Transmission curve of the \gal{} filter (green) taken from Gaia DR2 and the highly idealized transmission curve assumed for calculations in the \sj{} filter (red). The quiet-Sun irradiance spectrum (plotted in gray) used by the SATIRE model is shown in the background for reference.}
    \label{fig:filter}
\end{figure}

$\phi(\lambda)$ in \eqn{eq:flux} is the filter transmission profile. In this work, we consider the passbands of Gaia and Small-JASMINE missions. Gaia's astrometric instrument (ASTRO) measures the astrometric signal in the G-passband covering the visible and the near-infrared wavelength range of $330-1050$\,nm. Complementary to Gaia's measurements, \sj{} does astrometry in the infrared wavelengths in the range $1100-1700$\,nm. Figure~\ref{fig:filter} shows the transmission profile of the \gal{} passband taken from Gaia DR2\footnote{\url{https://www.cosmos.esa.int/web/gaia/iow_20180316}} and the assumed response curve of the \sj{} filter (the exact curve is not yet available in the literature). Note that the intensity in \eqn{eq:flux} is divided by the photon energy, ${hc/\lambda}$, because both the Gaia and \sj{} missions employ instruments measuring photon flux and not their energy. 

\begin{figure*}
\centering
\includegraphics[scale=0.6]{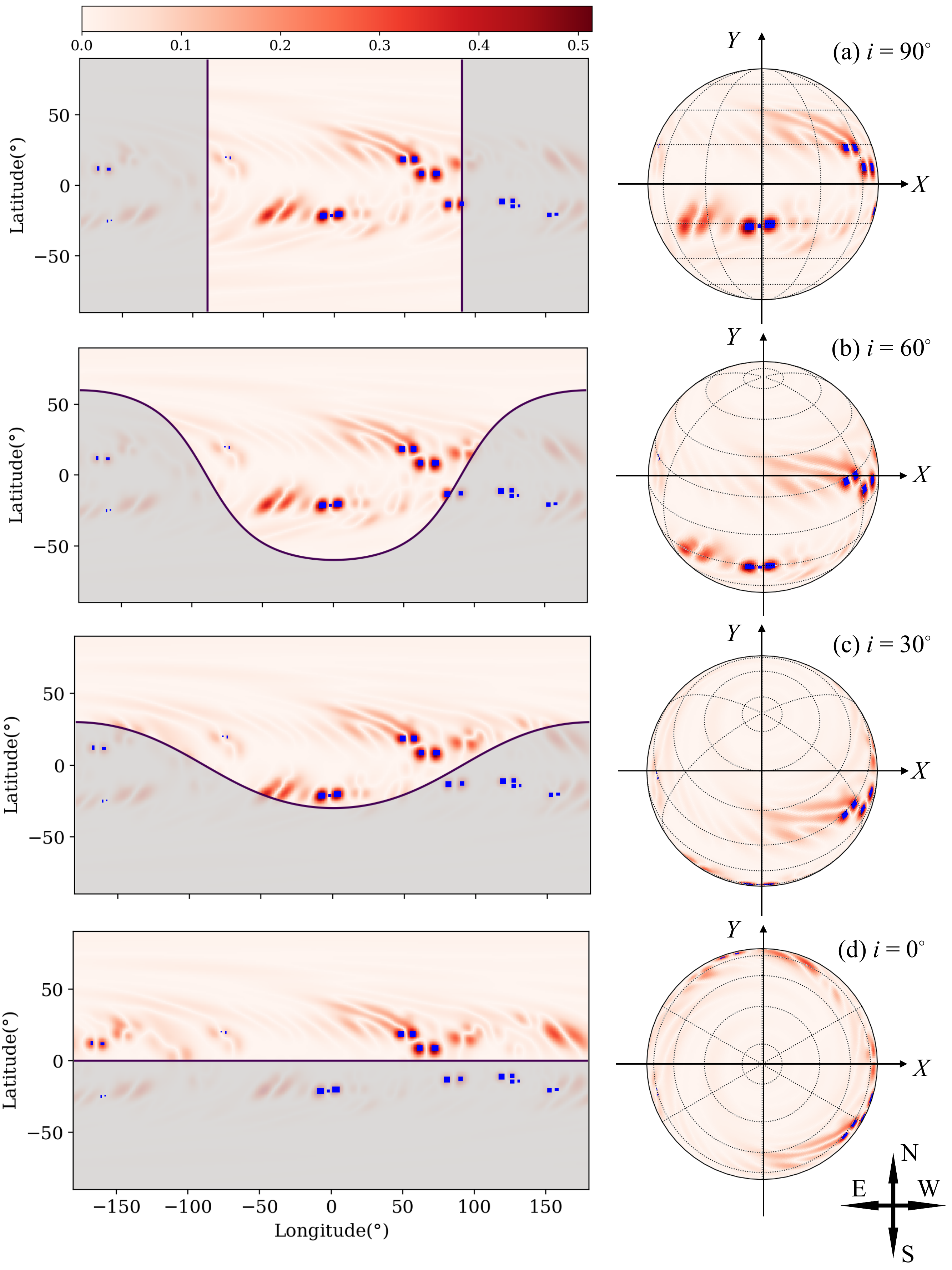}
\caption{Simulated distribution of spots (blue) and faculae (red shades) at a single time step around the maximum of activity cycle 22 for various inclinations as indicated at the upper right of the individual rows. The maps on the left show the distributions over the entire stellar surface. The shaded areas mark the far-side of the star. The right panels show the corresponding projected distributions on the visible stellar disk.}
\label{fig:sindis}
\end{figure*}

\begin{figure*}
    \centering
    \includegraphics[scale=0.6,angle=270,trim=3.5cm 0.0cm 3.0cm 0.0cm,clip]{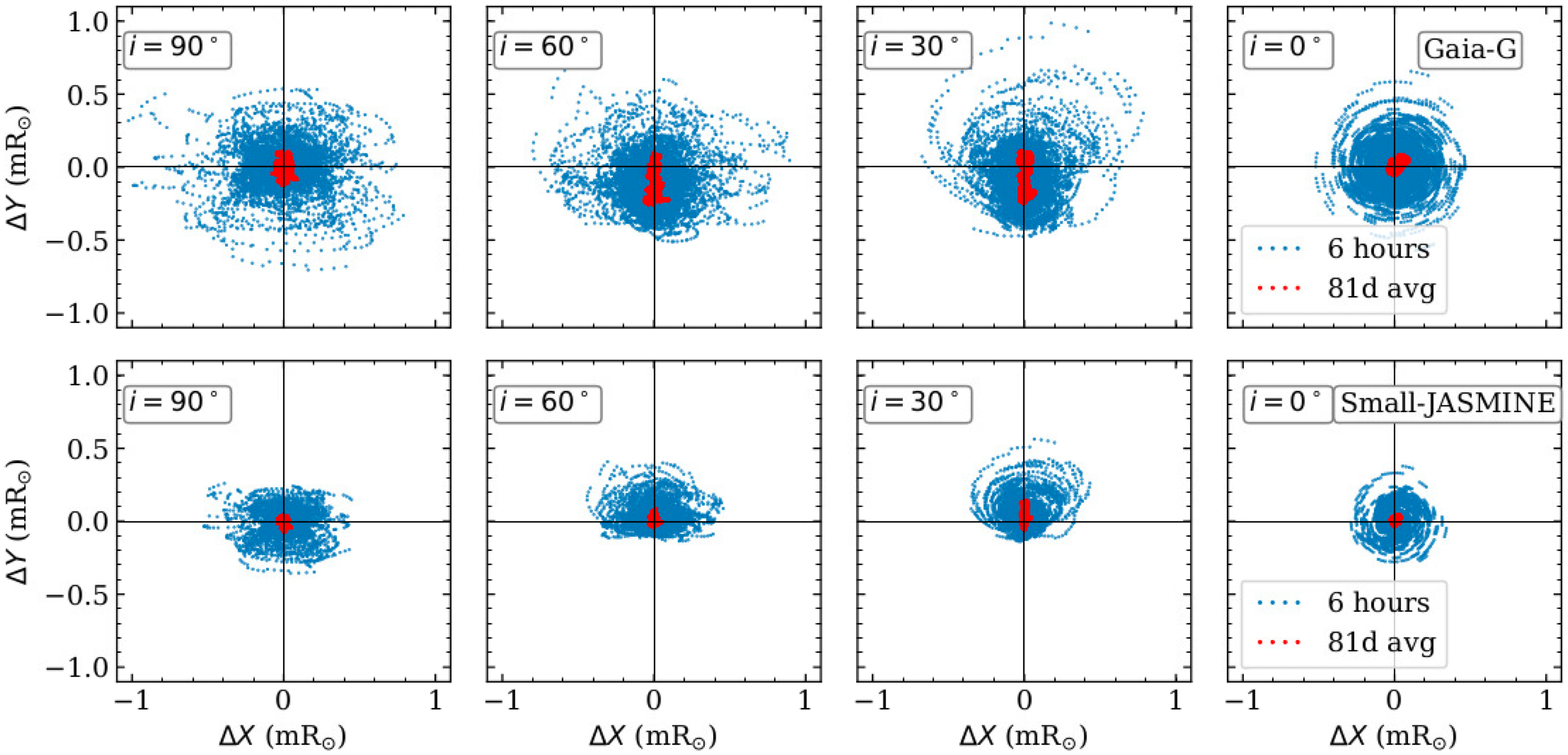}
    \caption{Displacements of the stellar photocenter as seen in \gal{} (upper panels) and \sj{} (lower panels) filters for different inclinations as indicated. $\Delta{X}$ and $\Delta{Y}$ are the shifts along $X$ and $Y$ directions, expressed in units of milli solar radii (mR$_\odot$). For an observer located 10\,pc away from the star, 1\,mR$_\odot$ corresponds to 0.5\,$\mu$as. Blue dots are the displacements calculated with 6-hour cadence and red dots are the 81-day moving averages.}
    \label{fig:jit}
\end{figure*}

\begin{figure*}
    \centering
    \includegraphics[scale=0.6,angle=270,trim=3.5cm 0.0cm 3.0cm 0.0cm,clip]{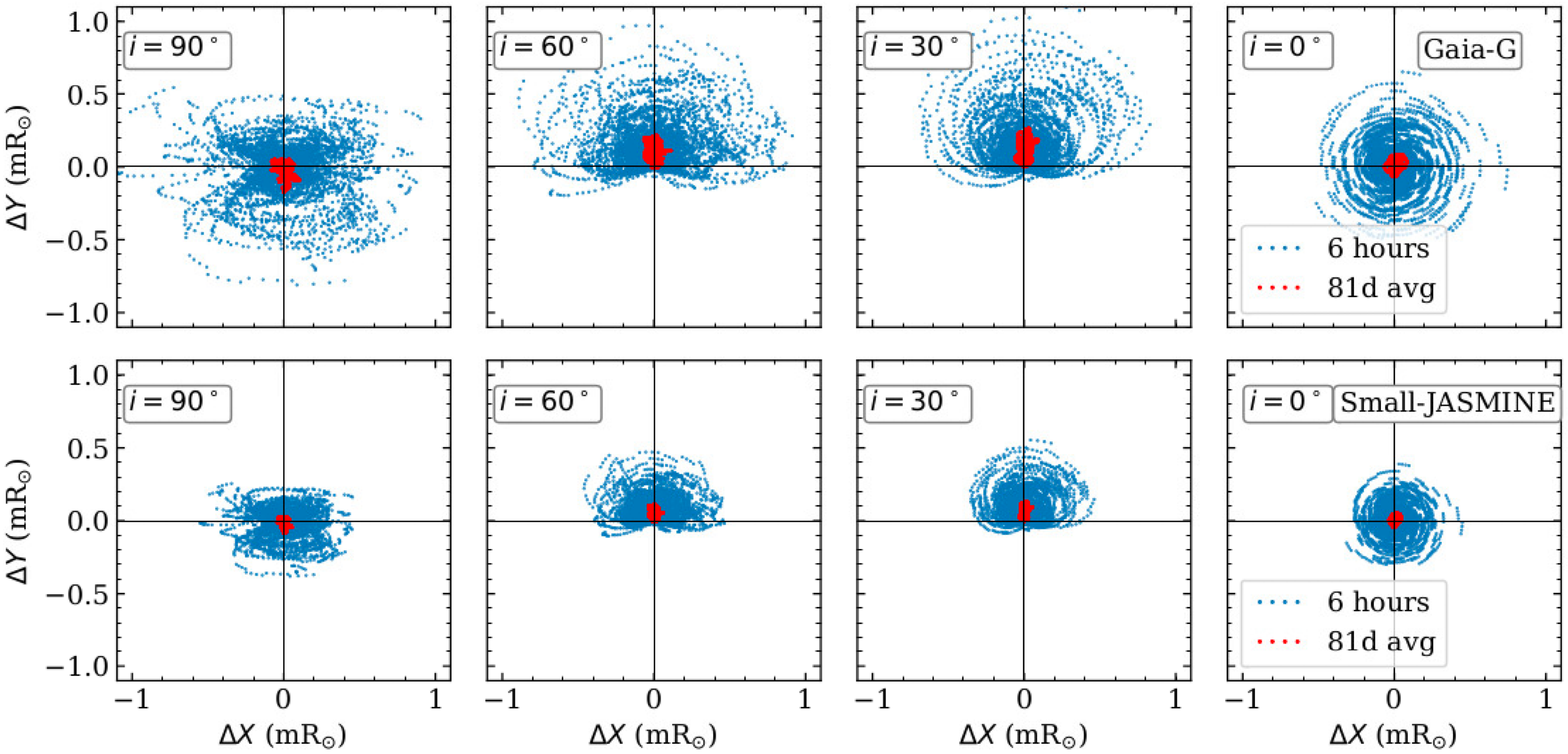}
    \caption{Similar to \fref{fig:jit} but only the contribution from spots is shown.}
    \label{fig:jitspot}
\end{figure*}

\begin{figure*}
    \centering
    \includegraphics[scale=0.6,angle=270,trim=3.5cm 0.0cm 3.0cm 0.0cm,clip]{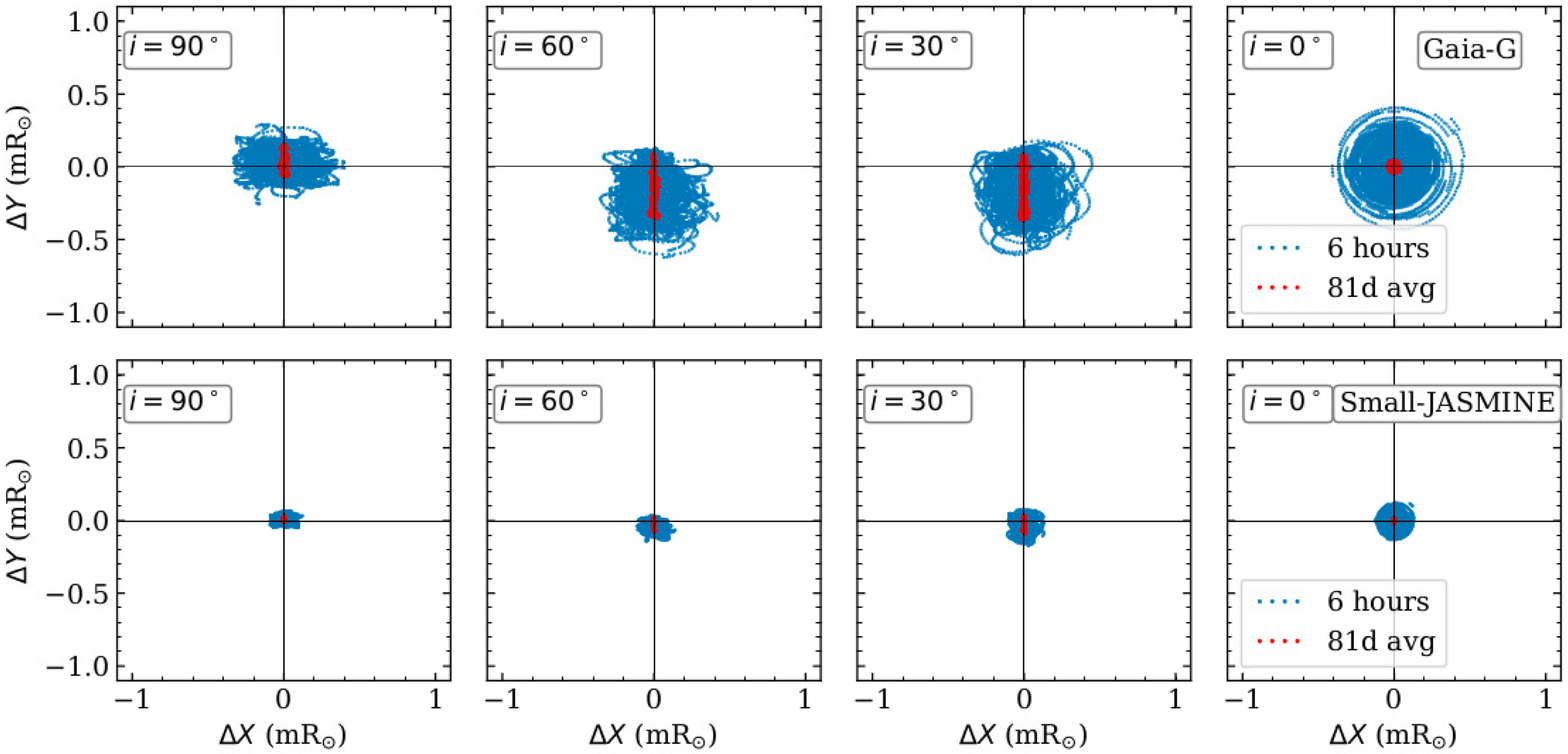}
    \caption{Similar to \fref{fig:jit} but only the contribution from faculae is shown.}
    \label{fig:jitfac}
\end{figure*}

\begin{figure*}
\centering
\includegraphics[scale=0.6]{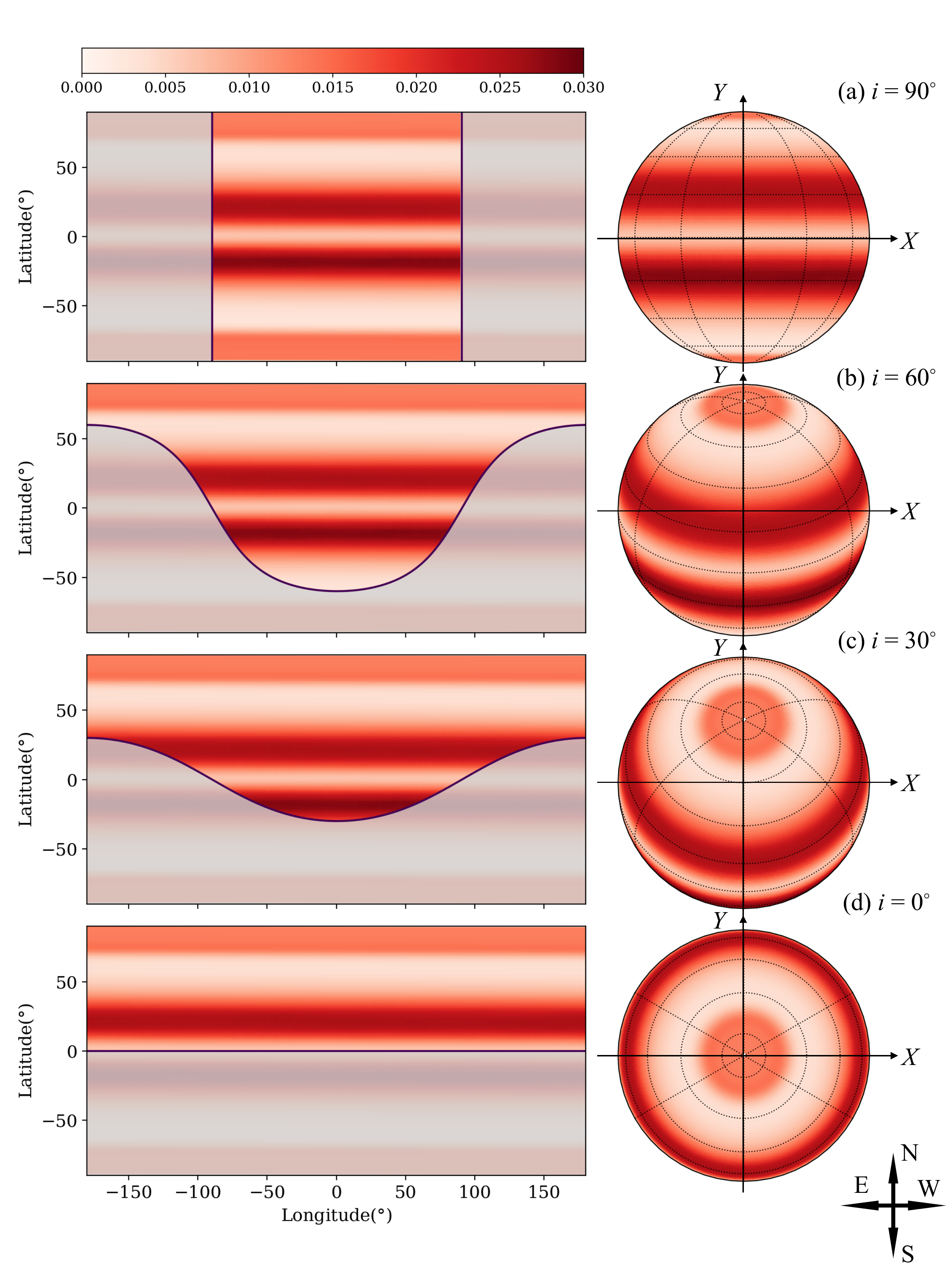}
\caption{Fractional area coverages of faculae averaged over cycle 22. The maps on the left show the distribution over the entire stellar surface. The shaded areas mark the far-side of the star. The right panels show the corresponding projected distributions on the visible stellar disk. Both the surface and the disk distributions have been calculated taking solar rotation into account. }
\label{fig:cycdis}
\end{figure*}

\begin{figure*}
\centering
\includegraphics[scale=0.6]{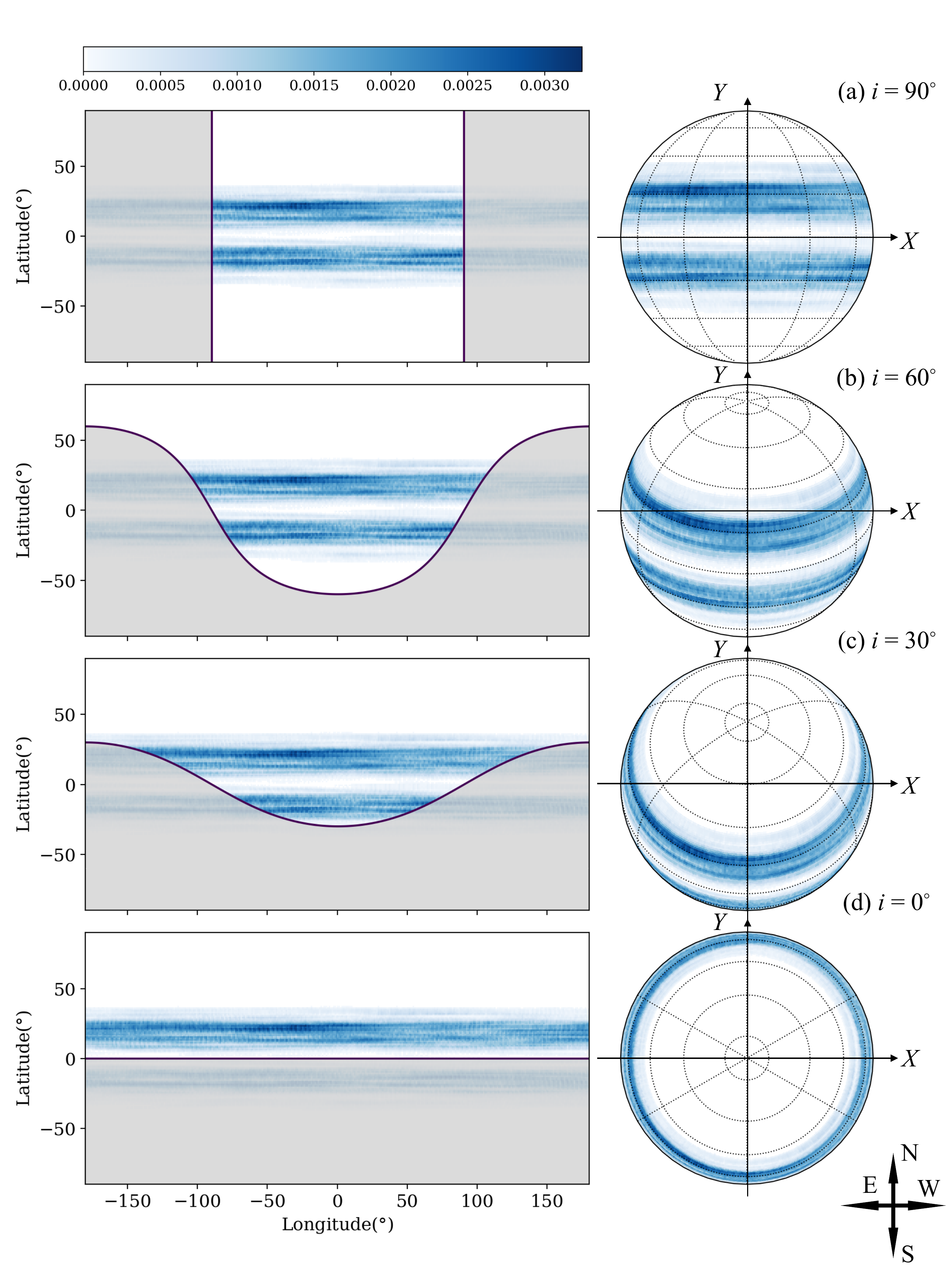}
\caption{Similar to \fref{fig:cycdis} but now for the fractional area coverages of spots.}
\label{fig:cycspot}
\end{figure*}

\begin{figure*}
    \centering
    \includegraphics[scale=0.58]{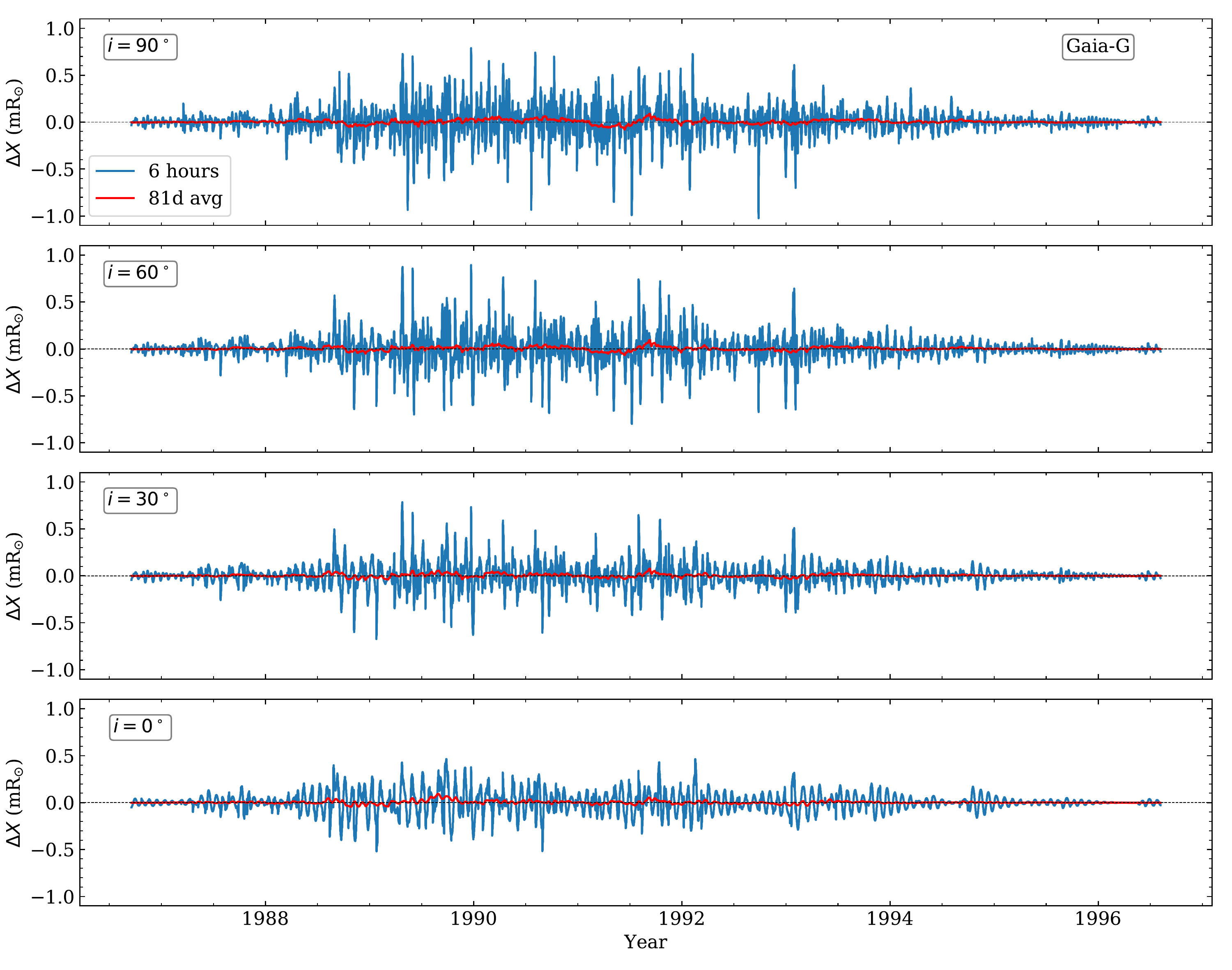}
    \caption{Time series of the displacement of the stellar photocenter in $X$ direction as seen in the \gal{} filter. The displacements with 6-hour cadence are shown in blue and the 81-day moving averages are shown in red.}
    \label{fig:xjit-gg}
\end{figure*}

\begin{figure*}
    \centering
    \includegraphics[scale=0.58]{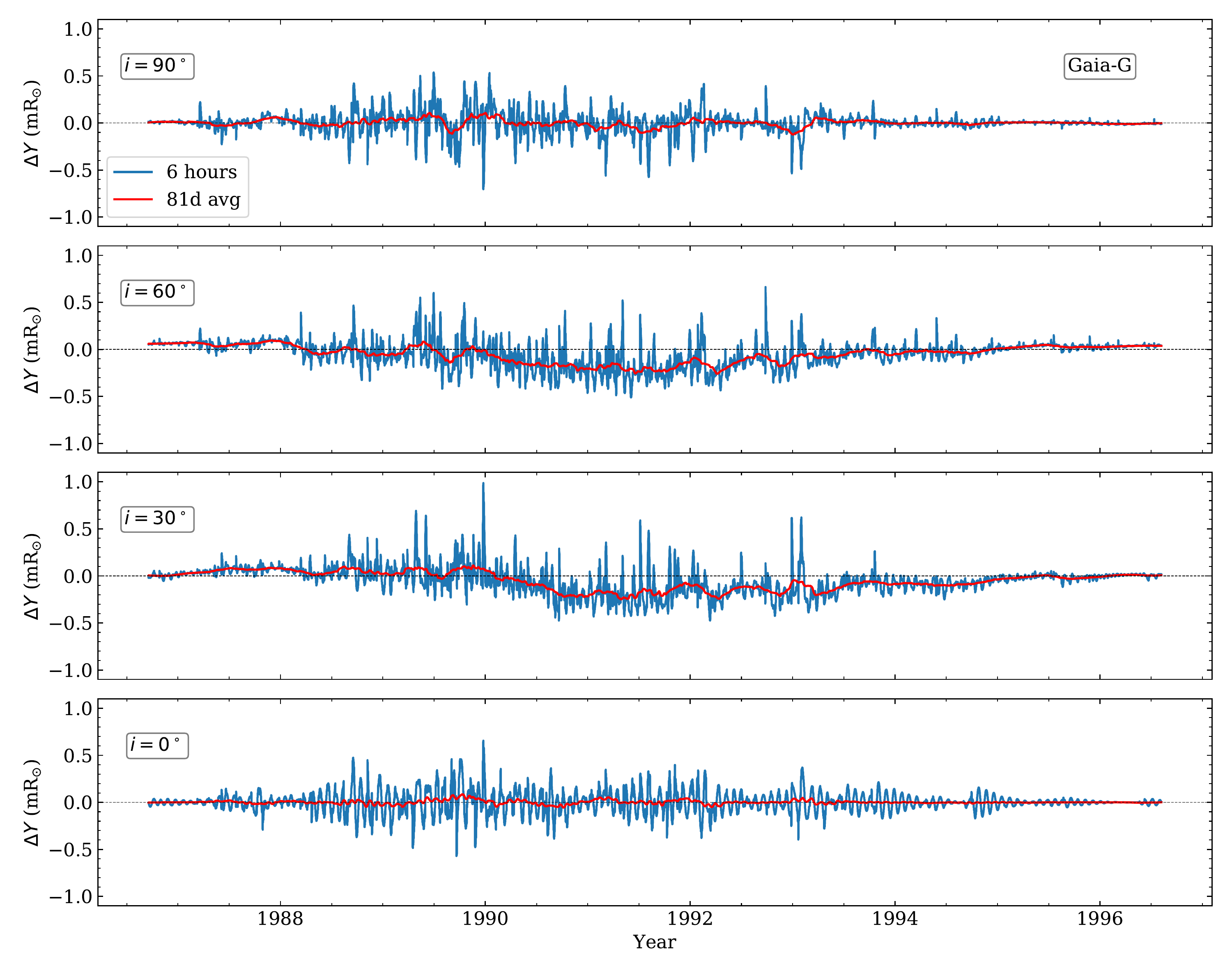}
    \caption{Similar to \fref{fig:xjit-gg} but for the displacement in $Y$ direction.}
    \label{fig:yjit-gg}
\end{figure*}

\section{Results}
\label{sec:res}
\subsection{Jitter on the timescale of stellar rotation}
\label{ssec:aj-rot}
As discussed earlier, the emergence of spots and faculae lead to the breaking of the spatial symmetry about the axis joining the visible disk center and the observer. On the Sun, spot and facular emergences are confined to low and intermediate latitudes. Figure~\ref{fig:sindis} shows a snapshot from the SFTM simulation and illustrates how the surface distribution of spots (blue) and faculae (red shades) transforms to the distribution on the visible disk for different inclinations. One can see from the figure that the distribution on the visible disk, which matters for the astrometric jitter, is clearly asymmetric about the origin (i.e. disk center). 

Due to this axially asymmetric distribution of the magnetic features, the photocenter gets displaced from the disk center, as shown by the blue dots in \fref{fig:jit}. The contributions of spots and faculae to this displacement are shown individually in Figures~\ref{fig:jitspot}~and~\ref{fig:jitfac}, respectively. The photocenter displacements are computed in both \gal{} and \sj{} filters for one complete activity cycle and are expressed in units of milli solar radii (mR$_\odot$). Daily displacements occur both along $X$ and $Y$ directions and the magnitude of the displacement changes with the inclination (see also \sref{ssec:char}). The displacements in the \sj{} filter are generally smaller than those in the \gal{} filter. This is because the contrasts of spots and faculae in the \sj{} filter are smaller than in \gal{} filter (see the blue curves in \fref{fig:cont}).

For an equatorial view (${i}=90\degree$), the distribution of the displacements is elongated in the $X$ direction due to the emergence of spots and faculae at low latitudes, and becomes more or less symmetric for the polar view (${i}=0\degree$). When viewed pole-on, the magnetic features appear near the limb (see panel d of \fref{fig:sindis}) and move along the limb as the star rotates. This affects the $X$ and $Y$ dimensions equally, thus leading to nearly symmetric displacements. The shape of the distribution gradually changes when the inclination changes. As shown in \fref{fig:sindis}, at ${i}=60\degree$ and ${i}=30\degree$, spots and faculae are concentrated mainly in the south half of the disk (i.e. at ${Y}<0$). Consequently, spots mainly displace the photocenter in the north direction (positive $Y$, see middle panels in \fref{fig:jitspot}). Faculae, however, predominantly lead to displacements in the negative $Y$ direction (see middle panels in \fref{fig:jitfac}). Since their latitude distribution is broader than that of spots, faculae also span some parts of the north half disk and lead to non-zero signal in the positive $Y$ direction. The trajectories of individual large spots transiting the visible disk are evident in \fref{fig:jitspot}.

\subsection{Jitter on the stellar activity cycle timescale}
\label{ssec:aj-act}
\sref{ssec:aj-rot} discusses the jitter on the rotational timescale. The behavior of the jitter changes on the timescale of the activity cycle. To illustrate this, we compute the 81-day smoothed values of the jitter (81 day corresponds to three times the mean solar rotation period of 27 days). The 81-day averaging reduces the displacements caused by the transit of magnetic features due to the solar rotation and shows the variations on the activity cycle timescale. The red points in Figures~\ref{fig:jit}--\ref{fig:jitfac} indicate such 81-day smoothed values. It is evident that the displacement in the $X$ direction is greatly reduced (compared to the daily displacement shown in blue). To understand this, we look at the cycle averaged distribution of faculae and spot area coverages. These are shown in Figures~\ref{fig:cycdis}~and~\ref{fig:cycspot}, respectively. The east (negative $X$ axis) and west (positive $X$ axis) parts of the visible disk are nearly symmetric for both spots and faculae. Thus the 81-day averaging lead to a reduction in the $X$ signal. In particular, faculae show almost no signal in the $X$ direction, in contrast to spots. The faculae have a longer lifetime than spots and produce more symmetric displacements as they transit the disk. The longer lifetime of faculae leads to a stronger averaging than in the case of spots. Therefore the 81-day averaged displacements in the $X$ direction in \fref{fig:jitfac} are smaller than those in \fref{fig:jitspot}.

Figures~\ref{fig:cycdis}~and~\ref{fig:cycspot} (left panels) also show that when averaged over a full activity cycle, the activity in the northern hemisphere is equal to the activity in the southern hemisphere. The resulting signal is therefore dictated by how the surface distribution transforms to the distributions on the visible disk. For ${i}=90\degree$, the north half of the visible disk corresponds to the northern hemisphere while the south part of the visible disk to the southern hemisphere. Both are equally active and hence the long-term displacement is zero. For inclinations $60\degree$ and $30\degree$, even though the surface distribution is symmetric, the symmetry is broken for the disk distribution since the northern activity belt gets projected mostly on to the southern half of the disk. For ${i}=0\degree$, the disk distribution becomes symmetric again and hence no signal is generated.

\begin{figure*}
    \centering
    \includegraphics[scale=1.0]{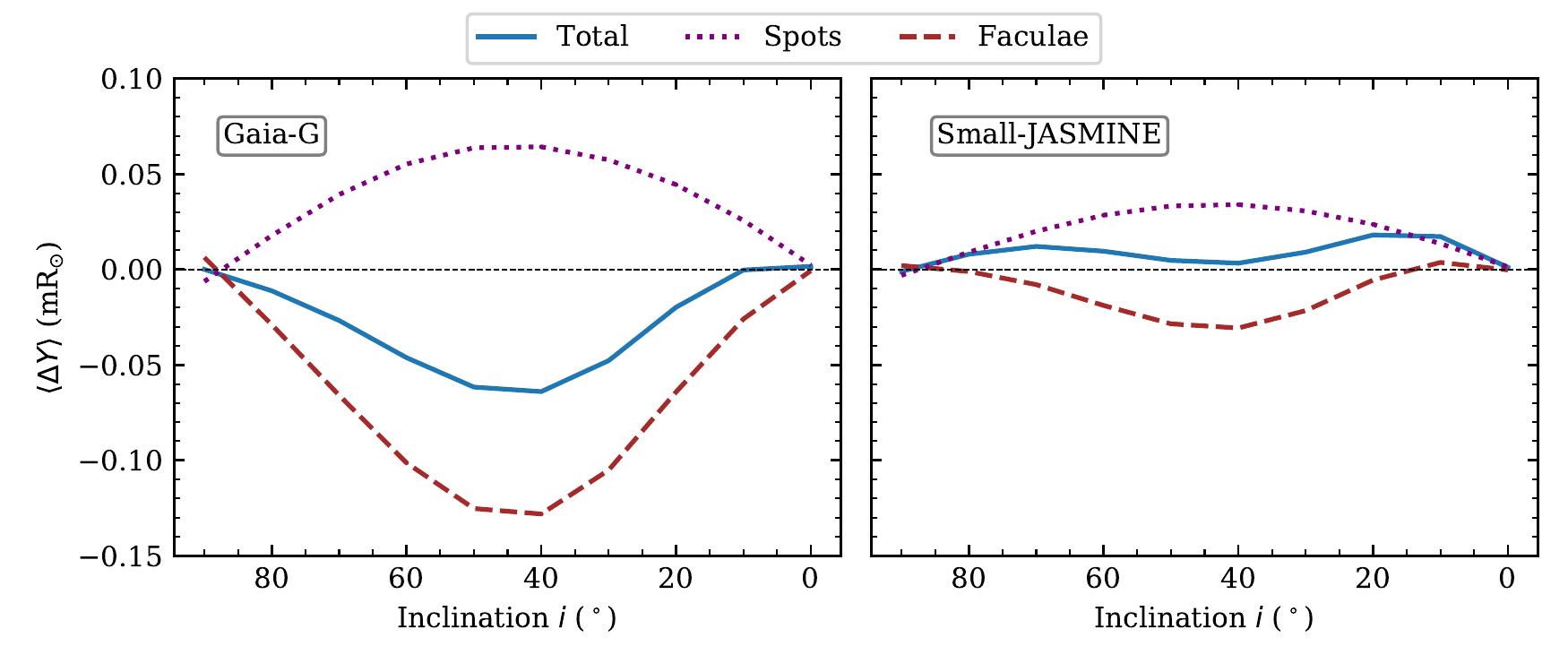}
    \caption{Cycle-averaged displacement of the stellar photocenter in $Y$ direction as seen in \gal{} (left) and \sj{} (right) filters as a function of inclination. The solid blue curve shows the total displacement, dotted purple and dashed brown curves show the average displacement caused by spots and faculae, respectively.}
    \label{fig:yjitavg}
\end{figure*}

\begin{figure*}
    \centering
    \includegraphics[scale=1.0]{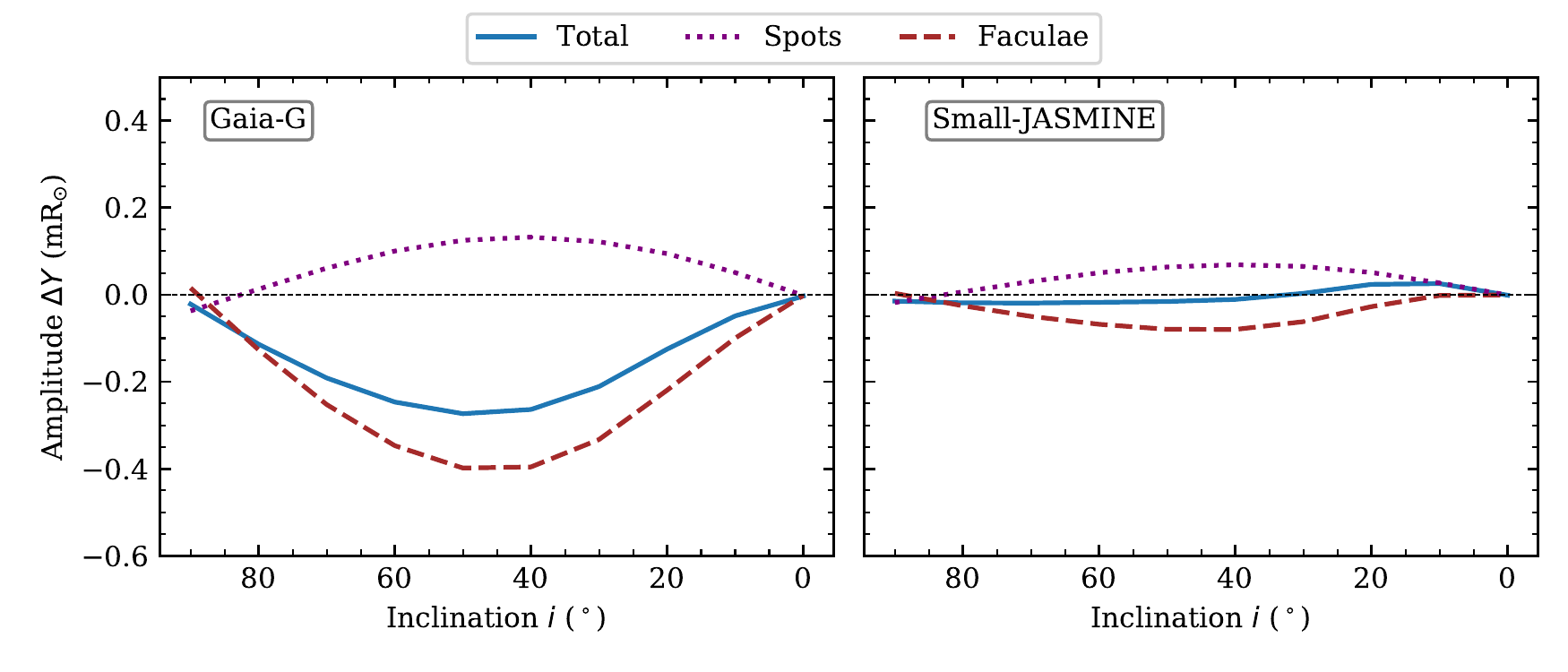}
    \caption{Amplitude of the photocenter displacement in $Y$ direction between the minimum and maximum of the activity cycle, as seen in \gal{} (left) and \sj{} (right) filters as a function of inclination. The solid blue curve shows the total displacement, dotted purple and dashed brown curves show the contributions from spots and faculae, respectively.}
    \label{fig:yjitamp}
\end{figure*}

\subsection{Time series of the displacements}
\label{ssec:char}
In the previous sections, we discussed the excursion of the photocenter in the $XY$ plane. In this section, we examine the time series of the astrometric jitter. Figures~\ref{fig:xjit-gg} and \ref{fig:yjit-gg} show the time variations of the photocenter displacement in the $X$ and $Y$ directions, respectively, in the \gal{} filter for various inclinations. The amplitude of the jitter on the rotational timescale is modulated by the solar activity during cycle 22. The photocenter displacements increase from the minimum to maximum of the activity cycle. The distinct individual spikes are due to transits of large spot groups on the visible disk.

Irrespective of the inclination, $\Delta{X}$ oscillates around zero (see \fref{fig:xjit-gg}). As a result, despite the vigorous rotational modulation, the 81-day averaged displacements in $X$ show very minor fluctuations, reflecting the east-west symmetry of the visible disk on the activity cycle timescale, as discussed in \sref{ssec:aj-act}. At the same time, for all inclinations (except for $90\degree$ and $0\degree$), the daily displacement in the $Y$ axis occurs about a non-zero mean level (see \fref{fig:yjit-gg}) showing a long-term variability, which is evident in the 81-day moving averages and is due to the breaking of the north-south symmetry (see discussion in \sref{ssec:aj-act}). Also, as the inclination changes, the polar regions become clearly visible in the north half of the disk (see Figure~\ref{fig:cycdis}). During the activity minimum, in the absence of spots, the magnetic flux in the north polar region leads to a shift of the photocenter above the equator (\fref{fig:yjit-gg}). As the activity cycle progresses, more spots and faculae emerge. The daily displacements are then determined by the interplay between the spots and faculae (see below). The time series of the displacements in the \sj{} filter have lower amplitudes and, in particular, have much lower values of the 81-day-averaged displacements (see Figures~\ref{fig:xjit-sj} and \ref{fig:yjit-sj}).

Since the displacements in $X$ average out due to the east-west disk symmetry, from here onward we consider only the displacements in $Y$. Figure~\ref{fig:yjitavg} shows the cycle-averaged displacements in $Y$ (corresponding to the mean position of the photocenter). The contributions of spots and faculae are also separately shown, in order to understand which magnetic feature dominates the cycle-averaged displacement for a given inclination. It is evident that the average shifts are at least ten times smaller than the shifts caused by individual transits of the magnetic features (compare the $Y$ axis in Figures~\ref{fig:yjit-gg}~and~\ref{fig:yjitavg}). However, it is interesting to note that the signal does not vanish for inclinations intermediate to equator-on and pole-on cases, i.e.~the mean position of the photocenter does not correspond to the visible disk center. Also, the mean shift as seen in the \gal{} filter is dominated by faculae whereas in the \sj{} filter mean shifts are dominated by spots at all inclinations. However, the mean shifts in the \sj{} filter remains smaller than those in the \gal{} filter.

Next, we compute the astrometric cycle amplitude as the difference between the time-averaged photocenter displacements along $Y$ in the periods $1991.3-1992.3$ and $1986.8-1987.8$. \fref{fig:yjitamp} shows the astrometric cycle amplitude in the \gal{} and \sj{} filters as a function of the inclination. In the \gal{} filter, the astrometric cycle amplitude increases with decreasing inclination, reaches a maximum value of 0.25\,mR$_\odot$ at ${i}=50\degree$ and then decreases to zero at ${i}=0\degree$. Also, the amplitude due to spots remains much lower than that due to faculae so that the astrometric cycle amplitude is dominated by faculae. In \sj{}, the spot and faculae contributions are comparable to each other at all inclinations but have different signs, leading to nearly zero astrometric cycle amplitudes.

\begin{figure*} 
    \centering
    \includegraphics[scale=0.58]{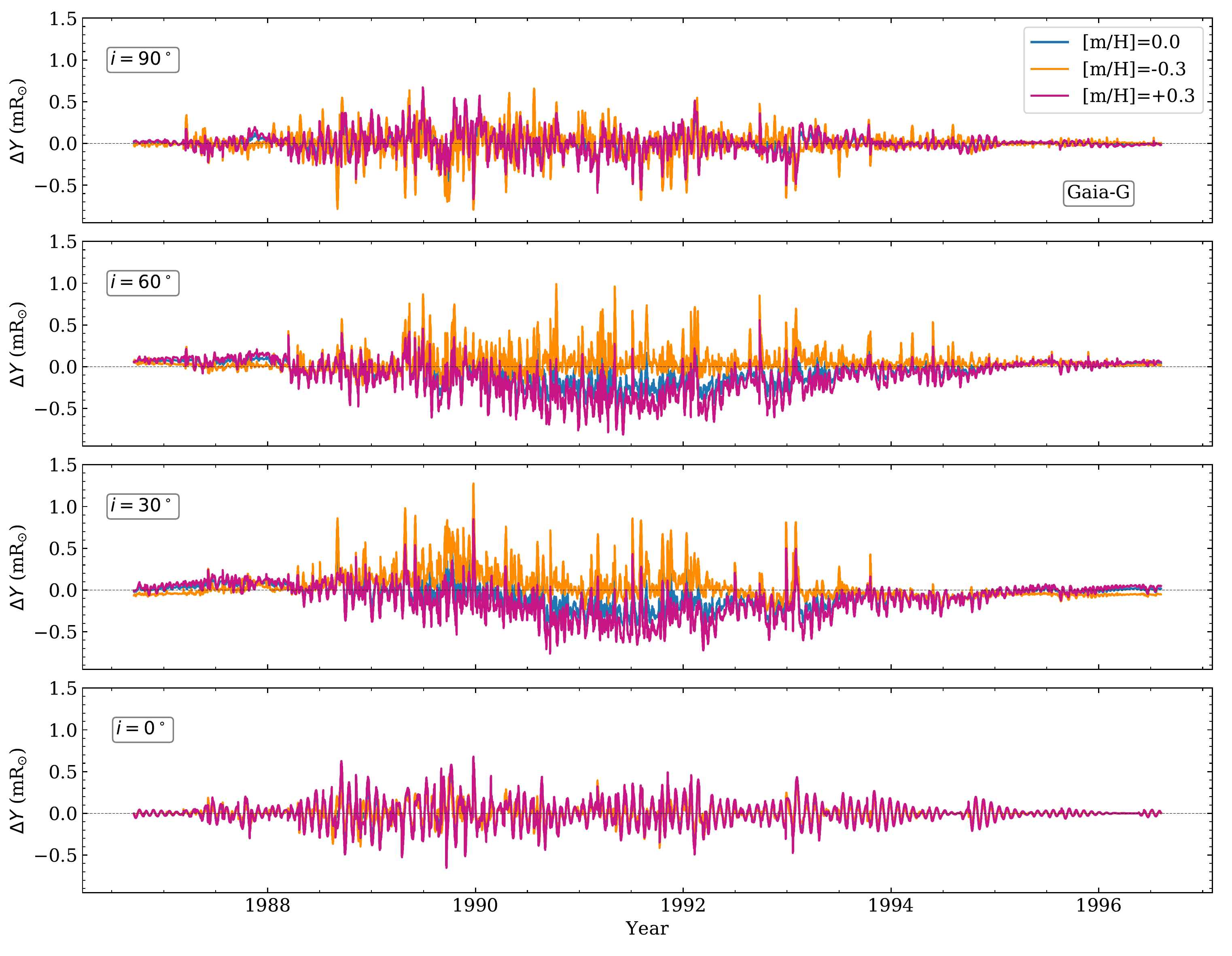}
    \caption{Effect of metallicity on the displacement of the photocenter in $Y$ direction as seen in \gal{} filter. Each panel represents the daily displacements for a given inclination as indicated. The case for solar metallicity ([m/H] = 0.0) is shown in blue, [m/H] = -0.3 and [m/H] = +0.3 are shown in orange and burgundy-red, respectively.}
    \label{fig:yjit-gg-met}
\end{figure*}

\begin{figure*}
    \centering
    \includegraphics[scale=0.58]{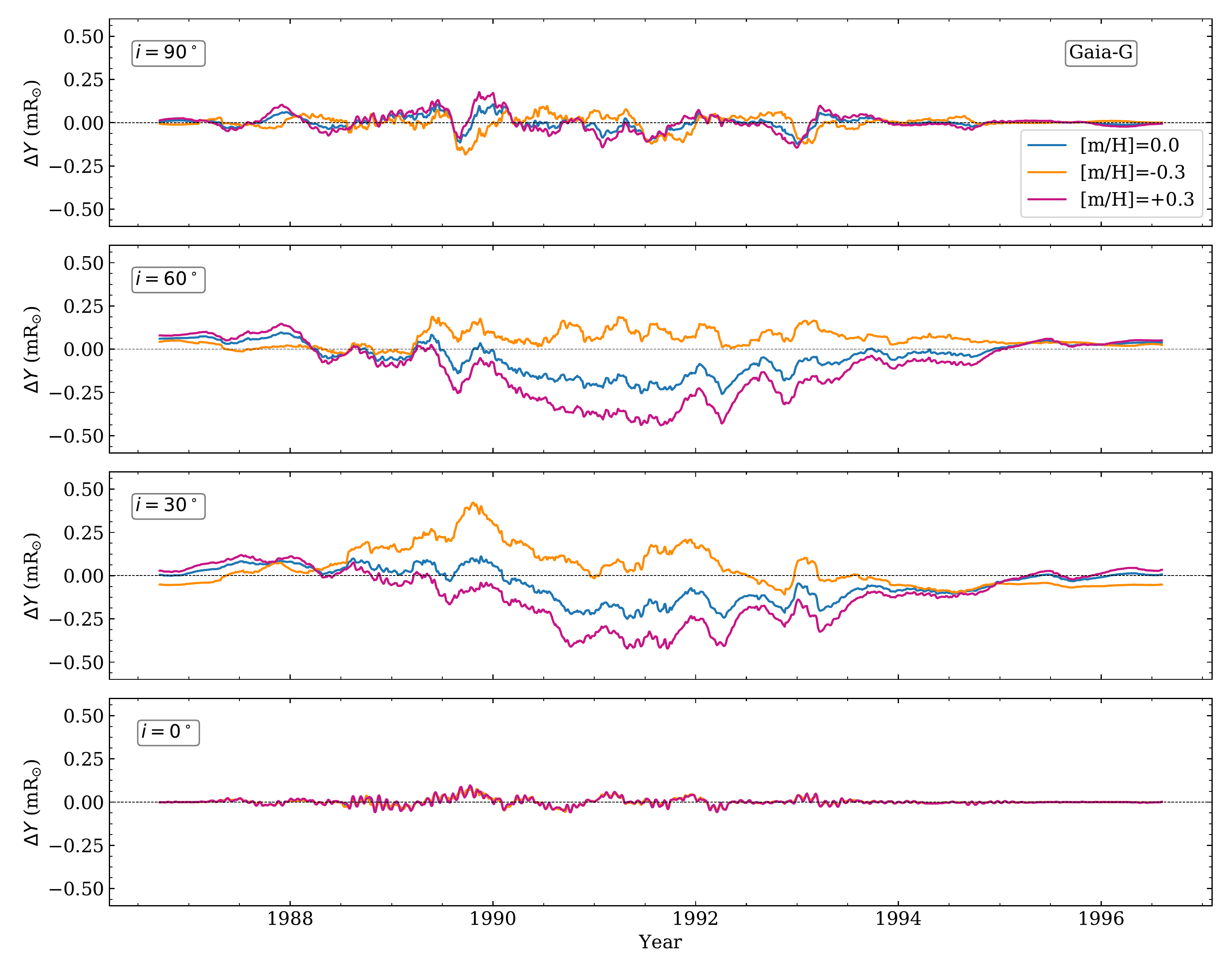}
    \caption{Same as \fref{fig:yjit-gg-met} but for the 81-day moving averages.}
    \label{fig:yjit-gg-met-81d}
\end{figure*}

\begin{figure*}
    \centering
    \includegraphics[scale=1.0]{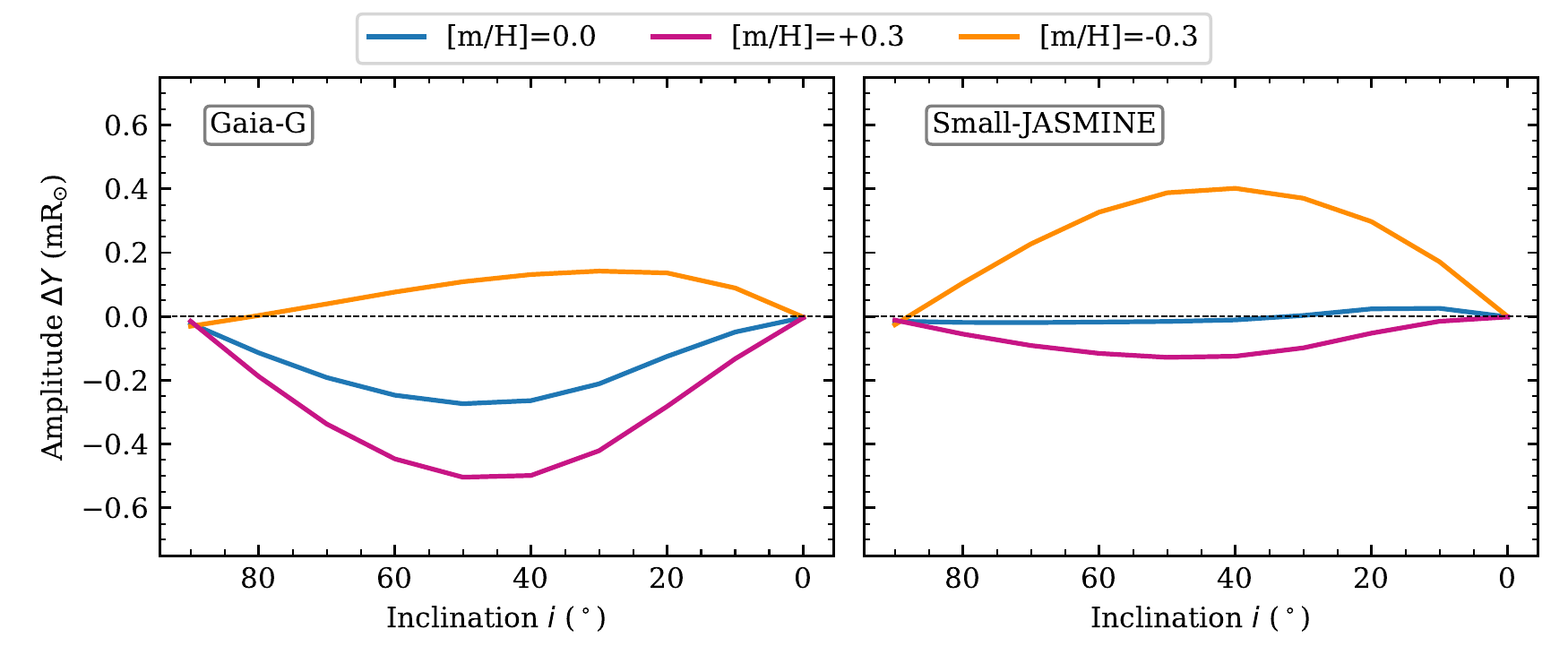}
    \caption{Amplitude of the photocenter displacement in $Y$ direction in \gal{} (left) and \sj{} (right) filters as a function of inclination for [m/H] = 0.0 (blue), [m/H] = -0.3 (orange), and [m/H] = +0.3 (burgundy-red).}
    \label{fig:yjitamp-met}
\end{figure*}

\subsection{Dependence on metallicity}
\label{ssec:met}
\cite{Veronika_2018} have shown that the brightness contrasts of the magnetic features depend on the metallicity, which is defined as
\begin{equation}
    [{\rm m/H}] = {\rm log}\bigg(\frac{N_{\rm metals}}{N_{\rm H}}\bigg)_{\rm star} - {\rm log}\bigg(\frac{N_{\rm metals}}{N_{\rm H}}\bigg)_\odot\ , 
\end{equation}
where $N_{\rm metals}$ is the number density of metals (all elements heavier than helium) and $N_{\rm H}$ is the number density of hydrogen. A comparison of the intensity contrast of the magnetic features with respect to the quiet-Sun computed by \cite{Veronika_2018} for the solar ([m/H] = 0.0), metal poor ([m/H] = -0.3) and metal-rich ([m/H] = +0.3) cases is shown in \fref{fig:cont}. One can see that the facular contrasts have a much stronger dependence on the metallicity than the spot contrasts \citep[see,][for a detailed discussion]{Veronika_2018}. 

The metallicity dependence of the facular and spot contrasts lead to a change in the amplitude of the astrometric jitter. This is illustrated in Figures~\ref{fig:yjit-gg-met} and \ref{fig:yjit-gg-met-81d} for the \gal{} filter (the daily values are plotted in \fref{fig:yjit-gg-met}, while the 81-day moving averages are shown in \fref{fig:yjit-gg-met-81d}). The displacements of the photocenter are not significantly different for the three metallicities when the star is observed equator-on and pole-on. In particular, for ${i}=0\degree$, the faculae and spots appear closer to the limb, where their contrasts are very similar for all three metallicity values.

For intermediate inclinations, however, the photocenter displacements show a strong dependence on metallicity. In particular, the variability on the activity cycle timescale is affected. During the cycle minimum period, the polar cap of faculae contribute to the displacement of the photocenter as previously mentioned. These polar regions appear in the north half of the disk and hence displace the photocenter towards them for [m/H] = 0.0 and [m/H] = +0.3. However, for [m/H] = -0.3, the facular contrasts become negative close to the disk center and hence the sign of the photocenter displacement caused by faculae is rather difficult to interpret. If they are closer to the disk center such as for ${i}=30\degree$, the faculae contrasts are negative (see \fref{fig:cont}) and hence repel the photocenter. Thus the displacements for [m/H] =  -0.3 are in a direction opposite to those of the other two metallicities for ${i}=30\degree$. During the rest of the cycle (i.e. periods beside minima), whenever the spots and faculae are closer to the disk center, facular contrasts add to those of spots and hence the photocenter gets repelled from the activity belt towards the equator. When the spots and faculae are away from the disk center, the facular and spot contrasts compete with one another leading to a reduction in the overall displacements.

Figure~\ref{fig:yjitamp-met} shows the dependence of the cycle amplitude of the photocenter displacement on metallicity and inclination. The amplitudes are computed following the description in \sref{ssec:char}. The separate contributions from spots and faculae for each case are shown in \fref{fig:yjitamp-gsj}. The dependence of the amplitude on inclination for [m/H] = +0.3 is similar to the solar case discussed in \sref{ssec:char} and the total amplitude is dominated by faculae at all inclinations in both \gal{} and \sj{} filters. The maximum amplitude is nearly twice that of the solar case in both filters. For [m/H] = -0.3, the spot and faculae contributions up to ${i}=50\degree$ as measured in the \gal{} filter act in opposite directions, almost canceling the astrometric signal whereas for inclinations between $50\degree$ and $0\degree$, the spot and faculae act in the same direction and thus enhance the signal. In the \sj{} filter, the amplitude of displacement has the same sign for both spots and faculae for all inclinations. This leads to a significant enhancement of the astrometric signal as compared to the solar case, where the signal was almost nonexistent. 

\begin{figure*}
    \centering
    \includegraphics[scale=0.7,angle=270,trim=0.0cm 2.2cm 0.0cm 2.0cm,clip]{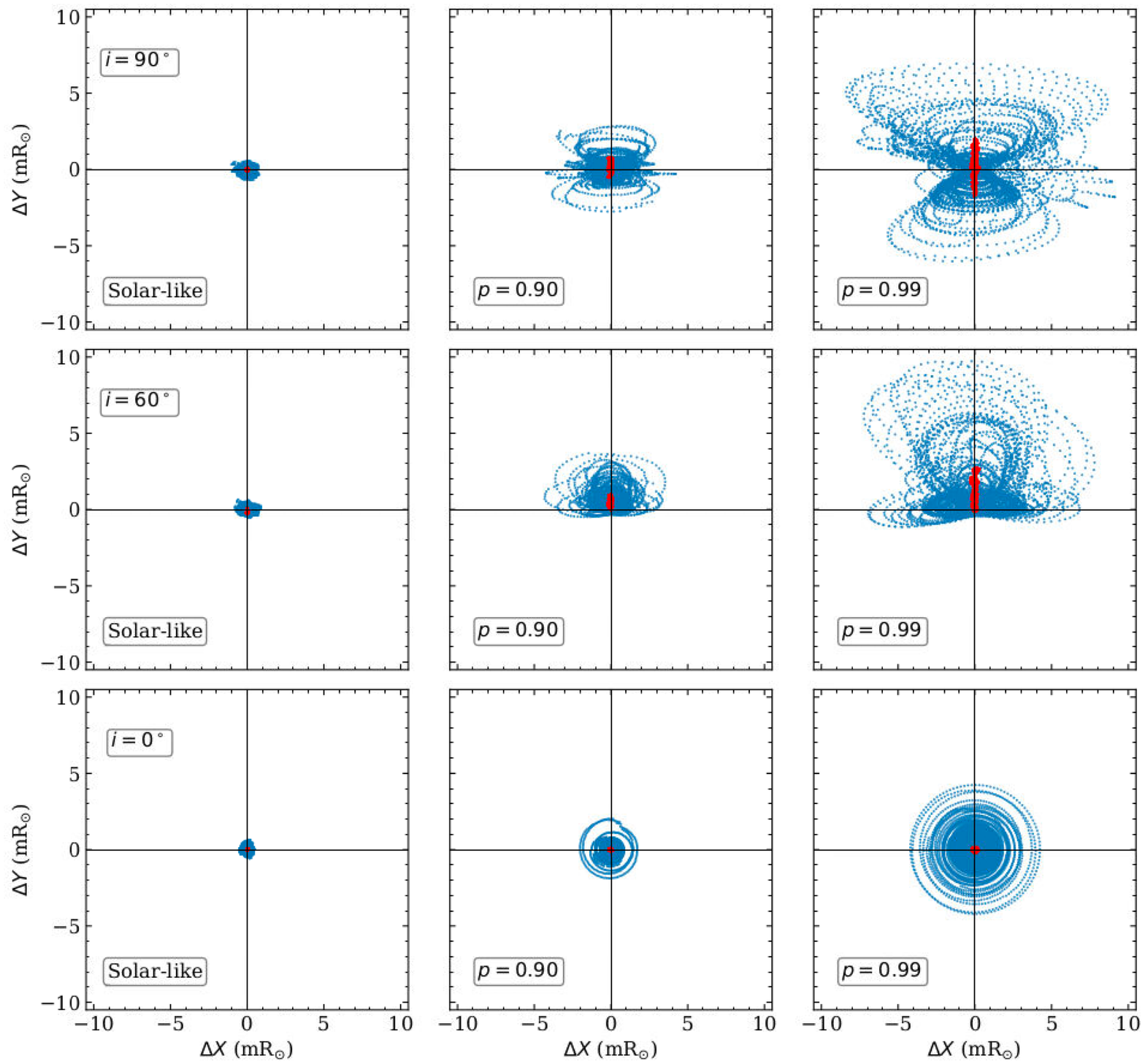}
    \caption{Photocenter displacements computed in the \gal{} filter assuming solar metallicity, for different active-region nesting probabilities ($p$) and inclinations. Left column: active region emergences as observed on the Sun, middle column: free-nesting with 90\,\%\ probability, right column: free-nesting with 99\,\%\ probability. Top row: ${i}=90\degree$, middle row: ${i}=60\degree$, bottom row: ${i}=0\degree$. Blue dots are the displacements computed at 6-hour intervals and red dots are 81-day moving averages.}
    \label{fig:nesting}
\end{figure*}

\subsection{Effect of active-region nesting}
The results presented so far have been obtained assuming a solar distribution of surface magnetic features. It is unclear if other stars with near-solar rotation periods and effective temperatures also display a similar distribution or not. Recently, \cite{Timo2020} found that a large group of stars with near solar-rotation periods and fundamental parameters observed by the \textit{Kepler} telescope have photometric variabilities substantially larger than that of the Sun. \cite{Emre2020} suggested that the large variability of these stars can be explained by a higher degree of active-region nesting on their surfaces than that observed on the Sun. A higher degree of nesting implies that these stars have surface magnetic features which are more inhomogeneously distributed than those on the Sun. Here, we calculate the possible values of astrometric jitter on such stars.

To model the inhomogeneous flux emergence caused by various degrees of active-region nesting, we employ the flux emergence and transport (FEAT) model presented in \citet{Emre2018}. While the FEAT model can synthesize the emergence patterns of active regions on stars with various levels of magnetic activity and rotation rates, in the present paper we only consider a star with a solar-level of magnetic activity and rotating with the solar rotation rate. \citet{Emre2020} described two nesting modes, namely the free-nesting and double active-longitude nesting. Here we only consider the free-nesting mode wherein the nests (compact groups of bipolar magnetic regions) occur at random locations within the activity belts \citep[see][for further details]{Emre2018}. Following \cite{Emre2018}, we denote the probability with which free-nesting occurs by $p$, where $0<p<1$. More specifically, $p$ is the probability that a given bipolar magnetic region is part of a nest.

To obtain the fractional area coverages of spots and faculae using the surface magnetic field maps from the FEAT model, we follow an approach that is different from \citetalias{Nina1}, which we used so far. \citet[in prep.; see also \citealt{Nina_thesis_arxiv}]{Nina2021} have proposed a masking procedure that is capable of accounting for possible formation of spots due to the superposition of small-scale magnetic flux. Such formation can happen when the magnetic flux density locally exceeds the threshold for the formation of spots, which can happen when multiple magnetic bipoles (i.e.~new active regions) appear at nearly the same position within a short period of time. Such overlapping emergences are a typical feature of large degrees of nesting (i.e. large values of $p$), so that we opted to follow the approach of \citet[in prep.]{Nina2021} in the calculations presented in this section. The spot masking method introduced by \citet[in prep.]{Nina2021} involves two magnetic flux density ($B$) thresholds, namely, $B_{\rm min}$ and $B_{\rm max}$, to determine the spot area coverages. The spot filling factor for a given pixel of the surface magnetic field map with $B\le B_{min}$ is assigned a value 0, whereas for a pixel with $B\ge B_{\rm max}$ the spot filling factor is given a value 1. For $B_{\rm min}< B < B_{\rm max}$, the spot filling factors are linearly interpolated between 0 and 1. The lower threshold prohibits the formation of polar spots (due to flux migrating to the poles) for slow rotators and an upper threshold acts as a saturation limit, above which a given pixel of the surface magnetic field map will be fully covered by spots. For computing the facular filling factors, \citet[in prep.]{Nina2021} follow the same means of setting a saturation threshold as done in \citetalias[and references therein]{Nina1}. The approach followed in \citet[in prep.]{Nina2021} results in the same level of solar rotational variability compared to the more complex \citetalias{Nina1} model, but is expected to yield more reliable results for stars exhibiting higher nesting than the Sun as well as for more active stars which will be considered in the next paper.

From Figure~\ref{fig:nesting}, it is evident that the jitter increases significantly with increasing nesting degree. Interestingly, the jitter for a star with a nesting degree of 90\,\%\ viewed pole-on exceeds the solar jitter at all inclinations, because the longitudinal inhomogeneity (the degree of non-axisymmetry) of magnetic flux becomes much larger than that on the Sun, leading to substantial displacements of the photocenter. Further, a star with a nesting degree of 99\,\%\ exhibits jitter which is substantially larger than that from the Sun. The peak-to-peak amplitude reaches over 10\,mR$_\odot$ (5$\mu$as for an observer situated 10\,pc away) for $p=0.99$. Such an amplitude of the jitter could be detected with continuous measurements from Gaia and could possibly interfere with the detection of planets in the habitable zone.

\section{Time series of the astrometric jitter as it would be obtained by Gaia}
\label{sec:det}
\begin{figure*}
    \centering
    \includegraphics[scale=1.0]{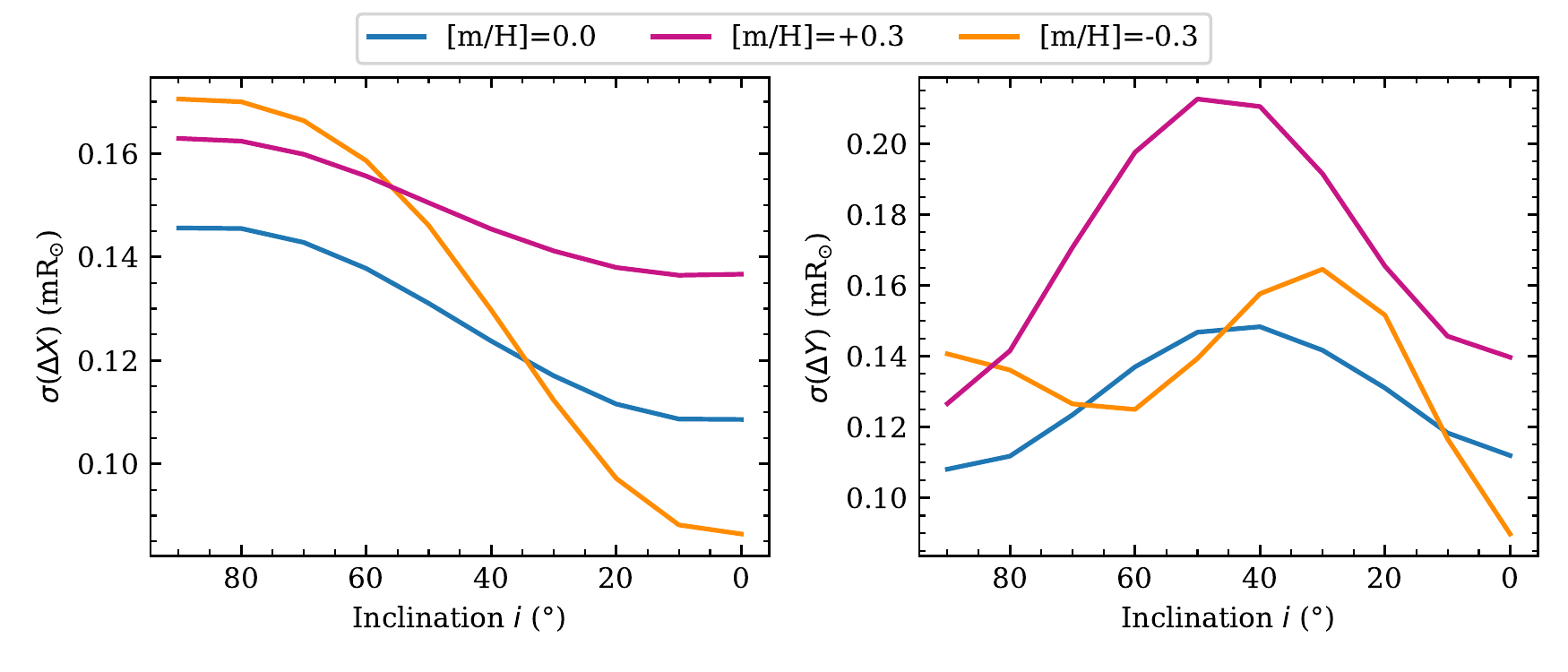}
    \caption{Standard deviation of the photocenter displacements in the \gal{} filter as a function of inclination, for different metallicities as indicated. Left panel: displacements in the $X$ direction; right panel: displacements in $Y$.}
    \label{fig:rmsxy}
\end{figure*}

\begin{figure*}
    \centering
    \includegraphics[scale=1.0]{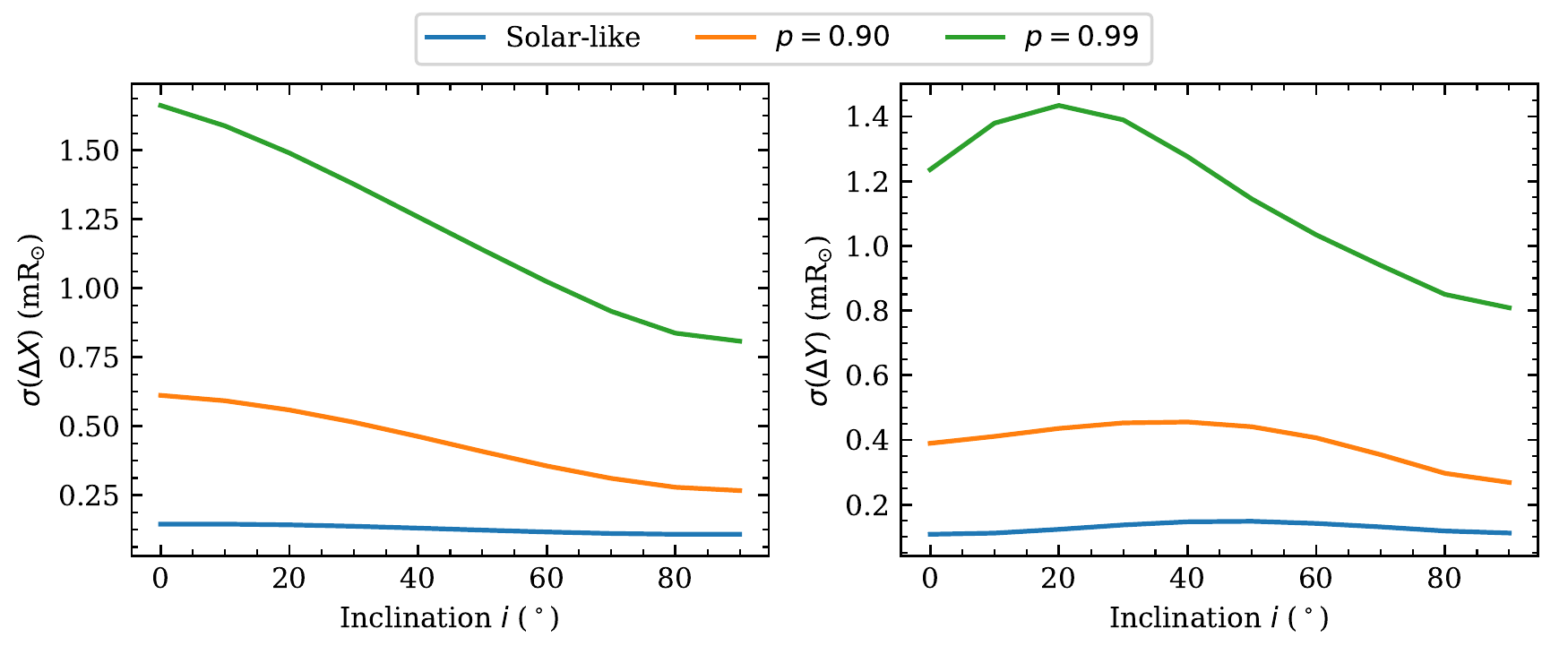}
    \caption{Standard deviation of the photocenter displacements in the \gal{} filter as a function of inclination, for different nesting degrees as indicated. Left panel: displacements in the $X$ direction; right panel: displacements in $Y$.}
    \label{fig:rms-nested}
\end{figure*}

Figures~\ref{fig:rmsxy}~and~\ref{fig:rms-nested} summarize the previously discussed results showing how the standard deviations of $\Delta{X}$ and $\Delta{Y}$ over the entire time interval considered in this study ($\sim 11$ years corresponding to solar cycle 22, see \sref{sec:model}) depend on the inclination, metallicity and degree of active-region nesting. We note that the plotted standard deviations are affected by the jitter both on the rotational timescale discussed in \sref{ssec:aj-rot} and on the activity timescale discussed in \sref{ssec:aj-act}. Although the deviations tend to increase with increasing metallicity and degree of nesting, the dependence on inclination is different for $\sigma(\Delta{X})$ and $\sigma(\Delta{Y})$, in accordance with \citealt{2020A&A...644A..77M}. 

Before simulating the astrometric jitter time series as it would be obtained by Gaia, we add the solar wobble caused by the Earth to the time series of daily displacements. The motivation behind considering the Sun-Earth system is to determine if the detection of Earth-mass planets in the habitable zones of stars with solar-like magnetic activity is influenced by the jitter due to magnetic activity of the host star. For the Sun-Jupiter system, the astrometric signal amplitude is three orders of magnitude larger than that for the Sun-Earth system. Such amplitudes are well above the precision offered by the current astrometry missions, so that the detection of Jupiter-mass planets is not affected by the magnetic activity of the host star.

It is rather straightforward to estimate that Earth going around the Sun at a distance of 1\,AU generates a wobble of 0.645\,mR$_{\odot}$. For an orbital period of one year, this wobble can be decomposed into sine and cosine waves with periodicities of one year and amplitudes of 0.645\,mR$_{\odot}$. Further, since the orbital plane of the Earth is nearly perpendicular to the solar rotation axis (neglecting the $7.1\degree$ tilt between the Earth's orbital plane and the Sun's equatorial plane), for an equator-on view, the Earth's motion does not generate any displacement of the photocenter in the $Y$ direction. On the other hand, for the pole-on view, equal displacements are seen in both $X$ and $Y$ directions. For intermediate inclinations, the $Y$ displacements get weighted by the angle between the orbital plane and the observer's line of sight.

\begin{figure*}
    \centering
    \includegraphics[scale=0.58]{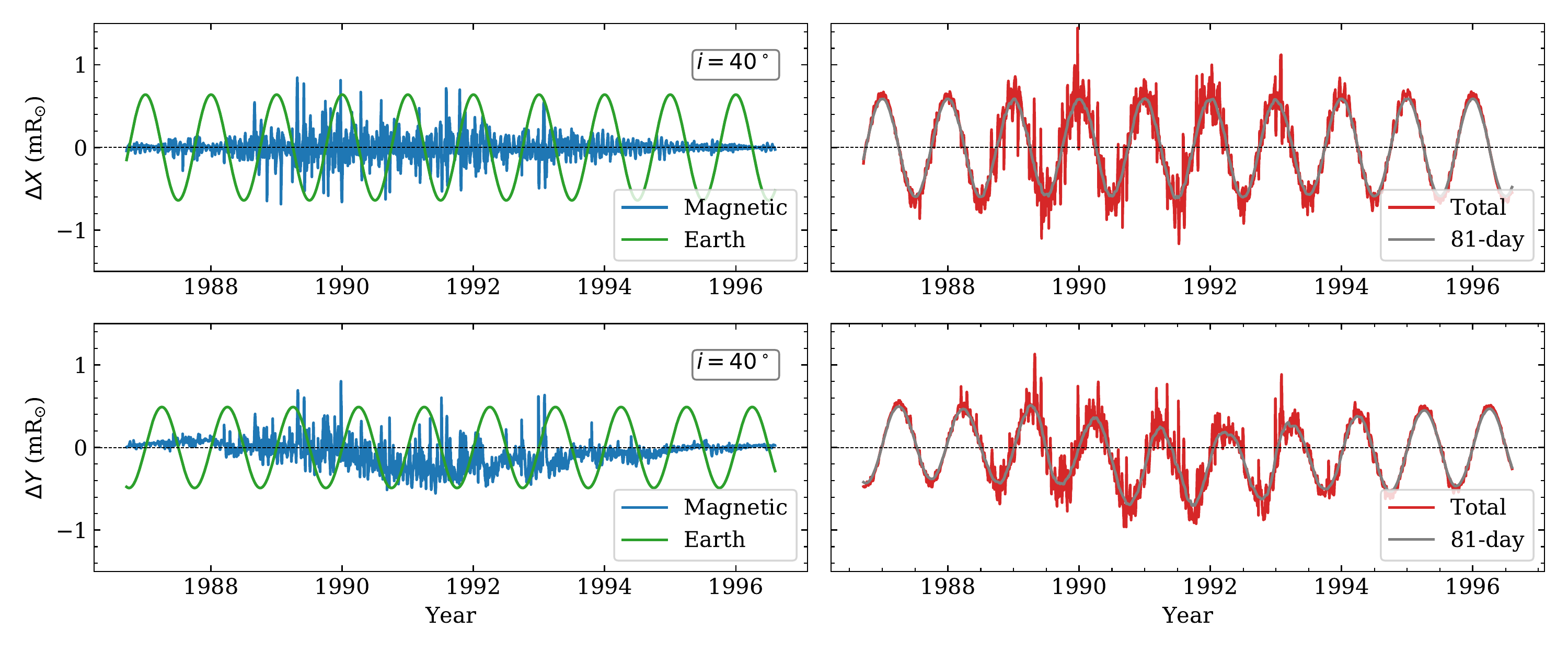}
    \caption{Photocenter displacements due to magnetic activity (blue) computed in the \gal{} filter, the astrometric signal due to Earth's motion around the Sun (green) and their sum (red). 81-day moving average of the total signal is shown in grey. Top panels show the displacements in $X$ directions while the bottom panels indicate the displacements along $Y$ direction.}
    \label{fig:stpl}
\end{figure*}

Figure~\ref{fig:stpl} shows the photocenter displacements in the $X$ and $Y$ directions, for an intermediate inclination of $40\degree$, caused by solar magnetic activity (in blue), the Earth (in green) and their sum (in red). It is clear from the figure (see left panels) that the peak-to-peak amplitude of the magnetic jitter is comparable to the astrometric signal introduced by the Earth's motion. In the total signal (i.e. $\Delta{X_{\rm total}}=\Delta{X_{\rm magnetic}}+\Delta{X_{\rm Earth}}$ and $\Delta{Y_{\rm total}}=\Delta{Y_{\rm magnetic}}+\Delta{Y_{\rm Earth}}$), during the minimum of the activity cycle when the jitter due to magnetic activity is low, the periodic variations due to Earth are hardly affected by the activity (see right panels). Even during periods of high activity, although the jitter amplitude increases, the periodic variations caused by the Earth are still clearly seen.

\begin{figure*}
    \centering
    \includegraphics[scale=0.58]{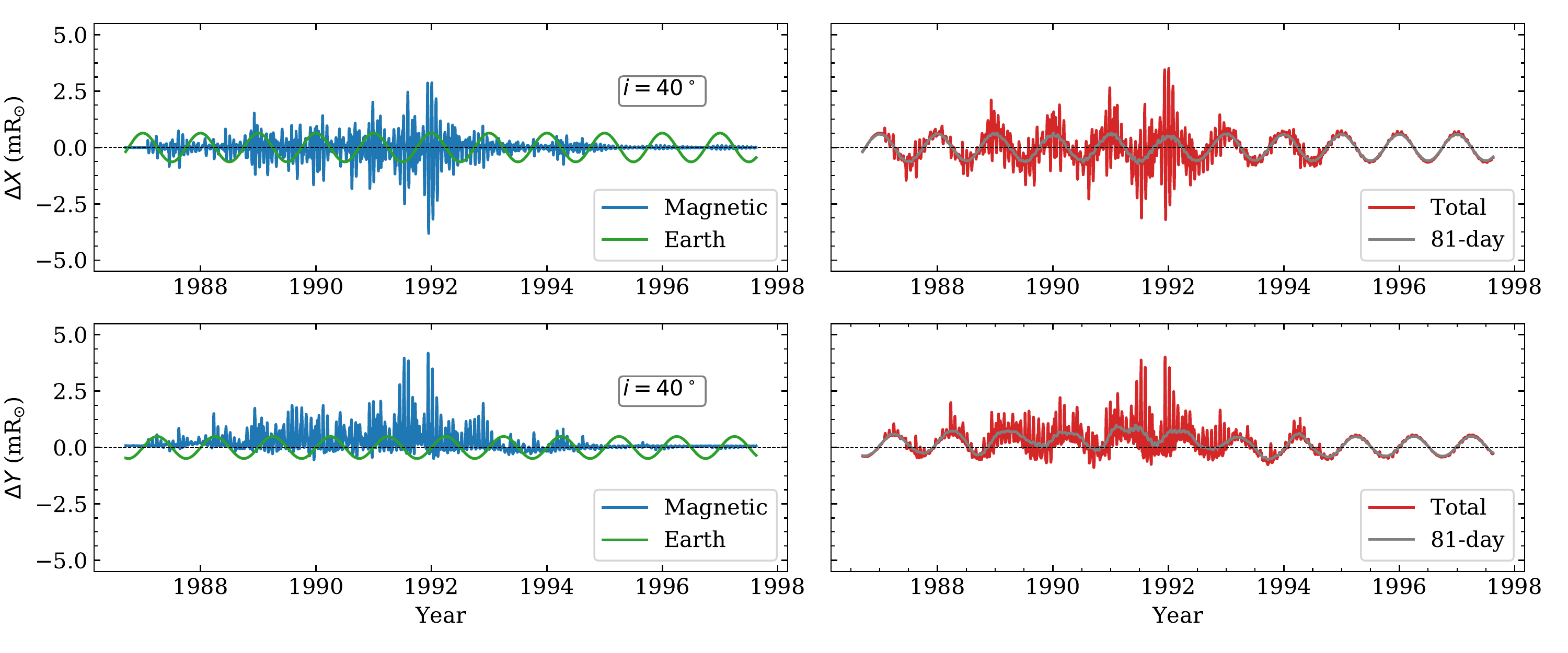}
    \caption{Same as \fref{fig:stpl} but now for the case of a nesting degree of 90\,\%.}
    \label{fig:stplnest}
\end{figure*}

When, however, the active regions emerge with $p=0.90$, the jitter due to magnetic activity is at least three times larger than that due to Earth, as shown in \fref{fig:stplnest}. As before, the Earth's signal remains distinct during the activity minimum whereas during the activity maximum, the magnetic activity dominates the jitter. Although the periodicity of the Earth's signal is almost preserved, the amplitude is reduced as shown by the 81-day moving averages (see grey curves in \fref{fig:stplnest}). If an instrument offers unlimited precision to measure such jitter amplitudes, then the characterization of an Earth-like planet can be significantly influenced by the stellar magnetic activity. 

Given these trends, the next task is to generate time series of the jitter as Gaia would see it. Gaia scans targets at non-uniform time intervals and records the displacements accurately only along the scan direction \citep[which changes with every visit of Gaia to a given star owing to the scanning law; see e.g.][]{2010IAUS..261..296L,2014ApJ...797...14P}. The time series of displacements, that Gaia would see, therefore depends both on the scan direction and the time intervals of observations. To illustrate how the time series of the displacement observed by Gaia would look, we use the observation strategy available at the Gaia events forecasts tool\footnote{\url{https://gaia.esac.esa.int/gost/index.jsp}} for GJ1243, which is a star with intermediate values of brightness ($G\sim11.5$\,mag) and ecliptic latitude ($\sim+46\degree$). This particular star is chosen owing to the dependence of the Gaia observation accuracy on the brightness and to the dependence of the number of Gaia visits to the star on the ecliptic latitude. The accuracy of Gaia's astrometric measurements is the highest for stars brighter than $G=11$\,mag and the number of Gaia visits to targets is largest at intermediate inclinations \citep[for more details, see][]{2014ApJ...797...14P}.

\begin{figure*}
    \centering
    \includegraphics[scale=0.58]{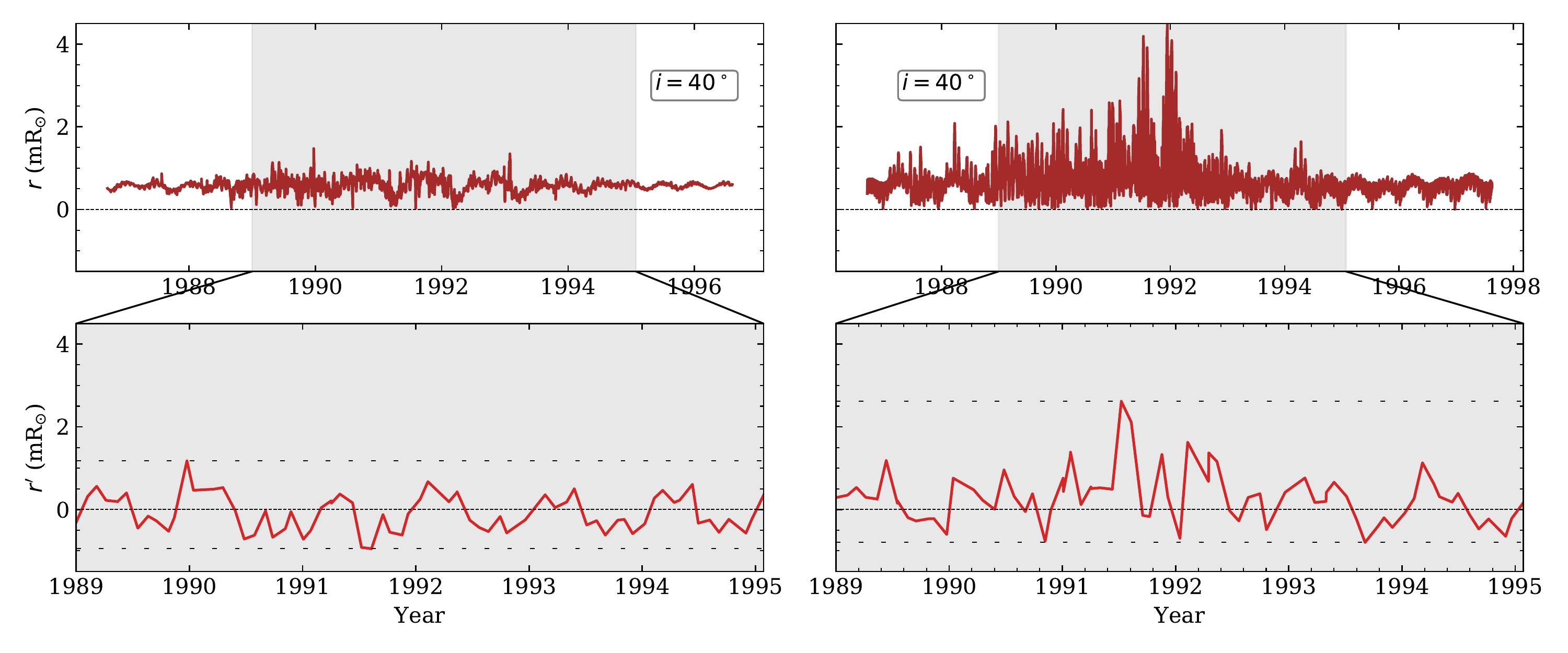}
    \caption{Absolute displacements of the photocenter for the Sun-Earth system (left panels) and for Earth-like planet going around a Sun-like star on which active regions emerge with a nesting degree of 90\,\%\ (right panels). The time series of daily displacements are shown in the top panels while the time series projected according to the observing strategy of Gaia are shown in the bottom panels. See text for more details.}
    \label{fig:gaia-ts}
\end{figure*}

The time series of daily displacements shown in red in Figures~\ref{fig:stpl}~and~\ref{fig:stplnest} are used to compute the absolute photocenter displacements from the true photocenter (disk center) as \mbox{${r}=\sqrt{\Delta{X_{\rm total}}^2+\Delta{Y_{\rm total}}^2}$}. The absolute displacements computed this way for the solar case and the case where active regions emerge with a nesting degree of 90\,\%\ are shown in the top panels of \fref{fig:gaia-ts}. We then interpolate the daily absolute displacements on to time intervals of Gaia visits, assuming that the first visit occurred on January 1, 1989. This generates the jitter time series as it would be measured when a star has an activity level the Sun had around the maximum of cycle 22. The displacements are then projected on to the scan direction of Gaia measurements. The projected absolute displacement, denoted as $r^\prime$, is assigned a positive sign if $\Delta{Y_{\rm total}}\ge0$ and a negative sign when $\Delta{Y_{\rm total}}<0$ \citep[see e.g.][]{2018MNRAS.476.5408M}. The time series of $r^\prime$ are shown in the bottom panels of \fref{fig:gaia-ts}. The peak-to-peak amplitude of the projected time series (indicated by the horizontal dashed lines) is about 2\,mR$_{\odot}$ ($\sim 1\mu$as at 10\,pc) for the solar case and about 3.5\,mR$_{\odot}$ ($\sim 1.75\mu$as at 10\,pc) for the nested case. It is interesting to note that the periodic signal due to Earth is no longer evident in the projected time series for the nested case.

\section{Conclusions and outlook}
\label{sec:concl}
With the advent of Gaia there has been a revived interest in investigating the possible interference of the magnetic-activity-induced jitter with the astrometric signal from the gravitational interaction in the star-planet system. In this study, we investigated the impact of inclination, metallicity and active-region nesting on the jitter as seen in the \gal{} and \sj{} passbands, considering stars with the solar rotation rate and solar-like butterfly diagram i.e., emergences of magnetic features are confined to low-latitude regions on either side of the equator. 

We found that on the activity-cycle timescale, the amplitude of the jitter in \gal{} filter increases with decreasing inclination, reaches a maximum around $50\degree$ and decreases again as the inclination decreases further. Although the jitter along $Y$ direction vanishes for inclinations of $90\degree$ and $0\degree$, owing to the north-south symmetry of the distribution of magnetic features on the activity-cycle timescale, it remains non-zero for intermediate inclinations. The jitter amplitude in the \sj{} filter remains minuscule at all inclinations. The changes in metallicity do not affect the amplitude of the jitter for equator-on and pole-on view. For all other inclinations, an increase in the metallicity leads to an increase in the jitter amplitude to nearly twice the value for solar metallicity. 

The daily displacements of the photocenter due to magnetic activity are found to be comparable to the astrometric signal produced by the Earth's orbital motion around the Sun. However, both the amplitude of the jitter and the signal from the Earth are sufficiently below the accuracy of single epoch astrometric measurement from Gaia. Interestingly, when the probability of an active region to be a part of an activity nest is increased (following the recent proposal by \citealt{Emre2020}, to explain the photometric variability of solar-like {\it Kepler} stars), the jitter is found to increase substantially, in extreme cases even to a level that could be detected with Gaia. 

The activity level of stars increases with increasing rotation rate, which is inversely proportional to age \citep[see e.g.][]{1972ApJ...171..565S}. As the activity level increases towards more active stars with stronger photometric variability, we expect the displacements of their photocenters to be larger than in the solar case, making them detectable with Gaia. In addition, the mean latitude of activity shifts towards the poles for faster rotators \citep[see e.g.][]{1992A&A...264L..13S,2001ApJ...551.1099S,2009A&ARv..17..251S}, another potential factor affecting astrometric jitter. Consequently, the next study in this series will focus on more active solar-type stars with shorter rotation periods. To obtain the visible distribution of magnetic features on such stars, we will employ the FEAT model \citep{Emre2018}, which calculates the emergence times, latitudes, and tilt angles of rising flux tubes, and simulates the surface flux transport process for a given mode of active-region nesting. 

\acknowledgments
We would like to thank an anonymous referee for their suggestions to improve the paper. K.S. received funding from the European Union's Horizon 2020 research and innovation programme under the Marie Sk{\l}odowska-Curie grant agreement No. 797715. N.-E.N., A.I.S. and V.W. have received funding from the European Research Council under the European Union's Horizon 2020 research and innovation program (grant agreement No. 715947). S.K.S. has received funding from the European Research Council under the European Union’s Horizon 2020 research and innovation programme (grant agreement No 695075) and has been supported by the BK21 plus program through the National Research Foundation funded by the Ministry of Education of Korea.

\appendix 
\restartappendixnumbering
\section{Additional figures}
\begin{figure*}
    \centering
    \includegraphics[scale=0.58]{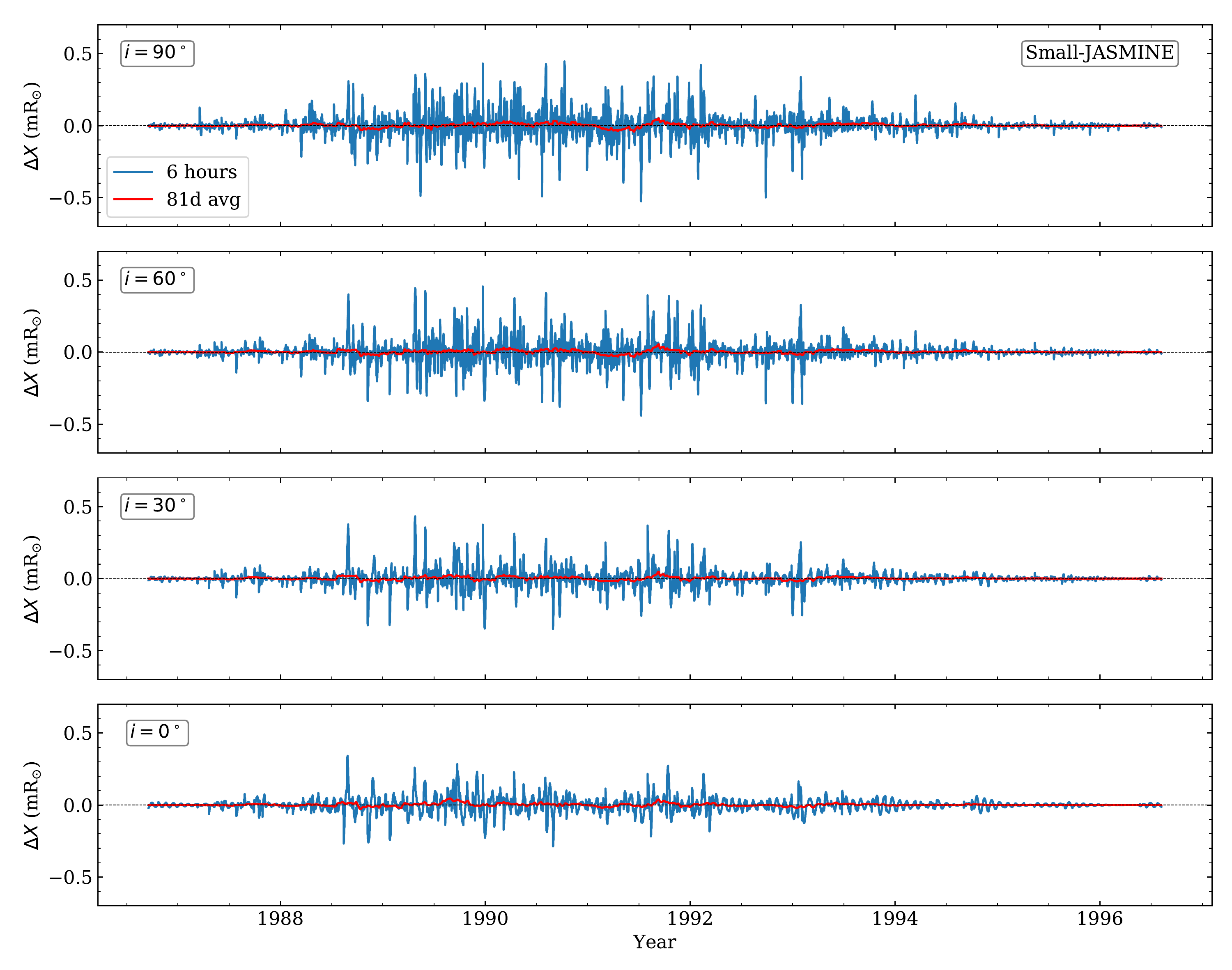}
    \caption{Time series of the displacement of the stellar photocenter in $X$ direction as seen in the \sj{} filter. Daily displacements are shown in blue and the 81-day moving averages are shown in red.}
    \label{fig:xjit-sj}
\end{figure*}

\begin{figure*}
    \centering
    \includegraphics[scale=0.58]{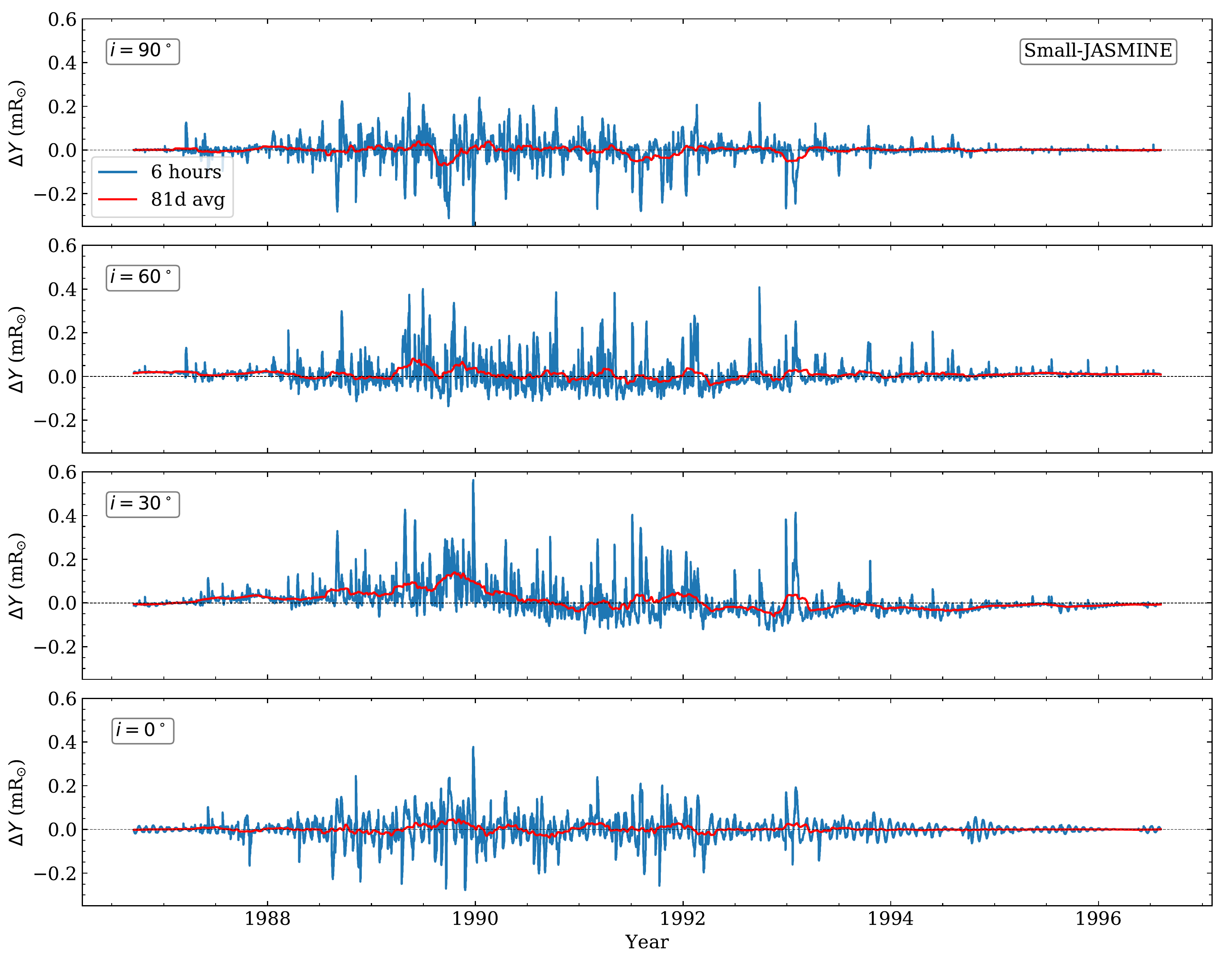}
    \caption{Time series of the displacement of the stellar photocenter in $Y$ direction as seen in the \sj{} filter. Daily displacements are shown in blue and the 81-day moving averages are shown in red.}
    \label{fig:yjit-sj}
\end{figure*}

\begin{figure*}
    \centering
    \includegraphics[scale=0.59]{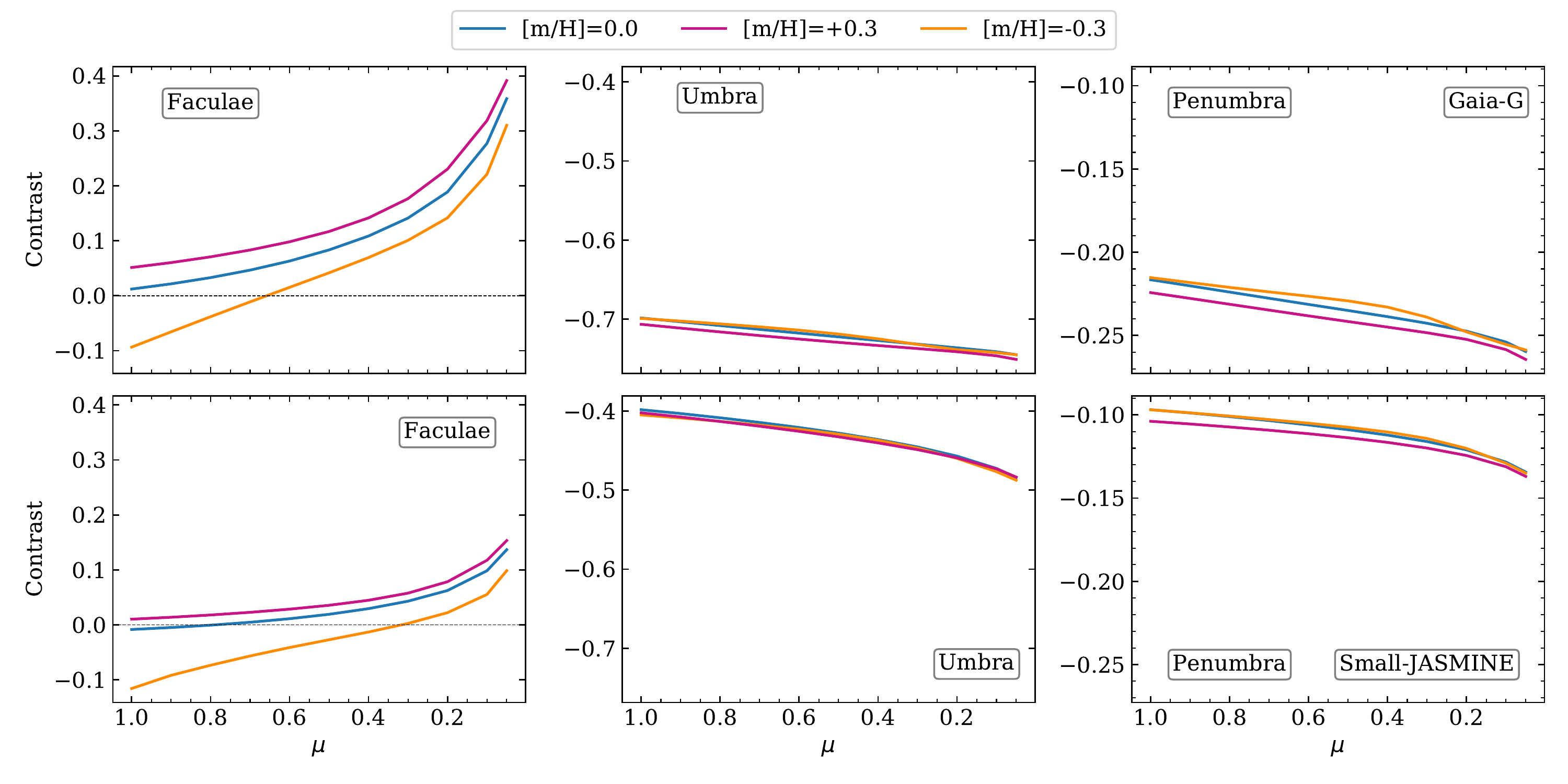}
    \caption{Intensity contrasts of faculae (left column), spot umbra (middle column), and spot penumbra (right column) with respect to the quiet photosphere for [m/H] = 0.0 (blue), [m/H] = -0.3 (orange), and [m/H] = +0.3 (burgundy-red). Contrasts in the \gal{} filter are shown in upper panels while those in the \sj{} filter are shown in lower panels.}
    \label{fig:cont}
\end{figure*}

\begin{figure*}
    \centering
    \includegraphics[scale=0.65]{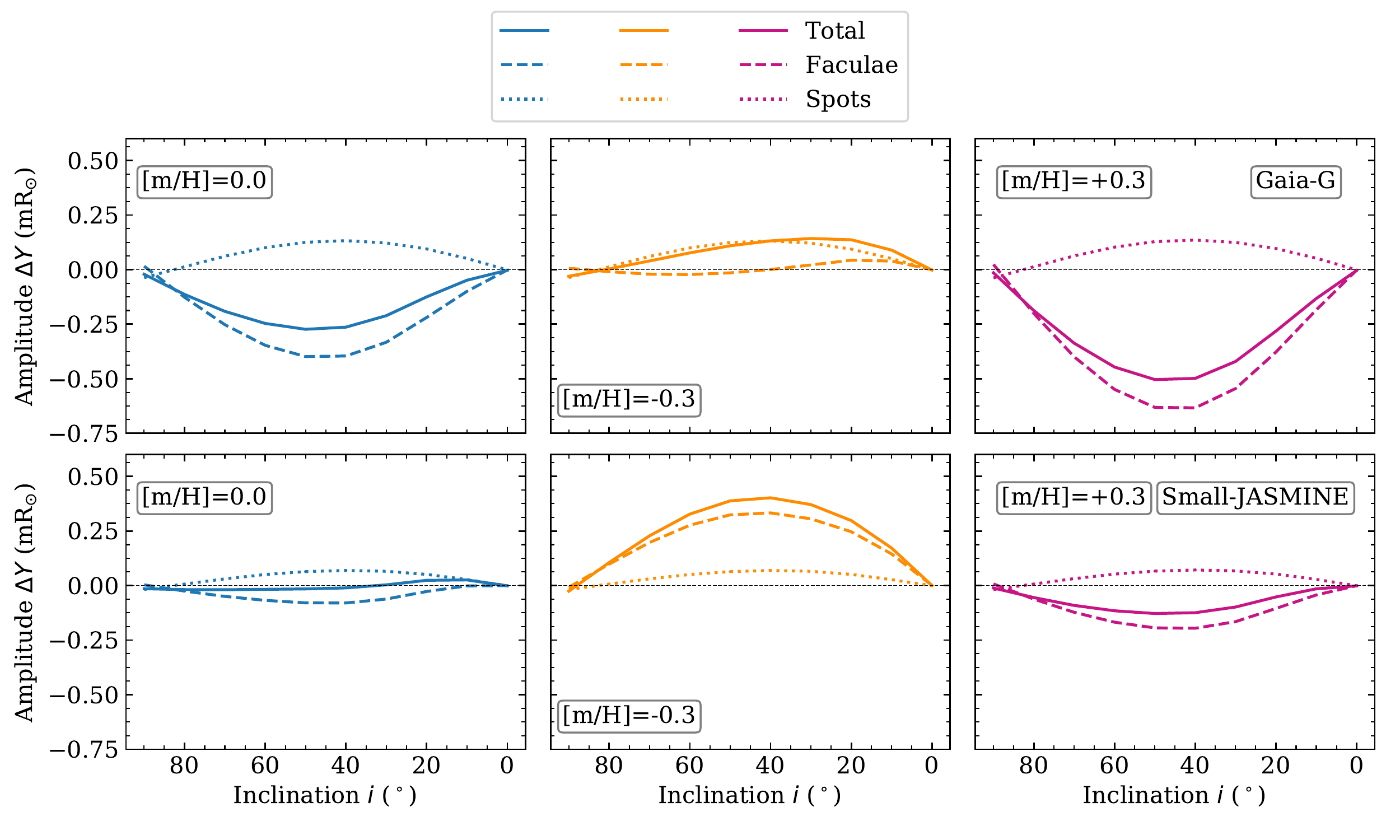}
    \caption{Amplitude of the photocenter displacement in $Y$ direction for [m/H] = 0.0 (blue), [m/H] = -0.3 (orange), and [m/H] = +0.3 (burgundy-red). Solid, dotted, and dashed lines show the total, spot, and facular contributions, respectively. Upper panels correspond to the \gal{} filter and lower panels to the \sj{} filter.}
    \label{fig:yjitamp-gsj}
\end{figure*}

Figures~\ref{fig:xjit-sj}~and~\ref{fig:yjit-sj} show the time series of the photocenter displacements along $X$ and $Y$ directions, respectively, as measured in the \sj{} filter. Figure~\ref{fig:cont} shows the center to limb variation of the intensity contrast of the magnetic features in \gal{} and \sj{} filters for different metallicities. The intensity contrast is defined as $[{I^k-I^0}]/{I^0}$. Figure~\ref{fig:yjitamp-gsj} shows the individual contributions of spots and faculae to the total astrometric cycle amplitude in \gal{} and \sj{} filters as a function of the inclination, for different metallicities.

\bibliography{astrojit}
\bibliographystyle{aasjournal}

\end{document}